# High-fidelity robotic PCR amplification

Vincent Beguin[1,*], Jean Grétillat[2,*], Kornelija Kaminskaitė[3], Simonas Juzenas[3], Dainius Kirsnauskas [3], Pierre-Yves Burgi[4], Samuel Wenger[1], Valentin Remonnay[1], Silvia Angeloni[1], Patrick Neuenschwander[1], Bart van der Schoot[5], Augustin Cerveaux[5], Thomas Heinis[6], Renaldas Raisutis[7], Martin Jost[8], Lukas Zemaitis[3], Ignas Galminas[3#], Jérôme Charmet[1,2,9#]

1. *School of Engineering−HE-Arc Ingénierie, HES-SO University of Applied Sciences Western Switzerland, Neuchâtel, Switzerland*
2. *School of Precision and Biomedical Engineering, University of Bern, Bern, Switzerland*
3. *Department of DNA data storage, Genomika, Kaunas, Lithuania*
4. *IT Department, University of Geneva, Geneva, Switzerland*
5. *Seyonic SA, Neuchâtel, Switzerland*
6. *Department of Computing, Imperial College London, London, United Kingdom*
7. *Ultrasound Research Institute, Kaunas University of Technology, Kaunas, Lithuania*
8. *MABEAL GmbH, Graz, Austria*
9. *Warwick Medical School, University of Warwick, Coventry, United Kingdom*

* the authors contributed equally to the work.

[#] corresponding authors

## Abstract

Polymerase chain reaction (PCR) underpins modern molecular biology, yet its deployment in emerging domains such as DNA data storage and distributed diagnostics remains constrained by bulky thermocyclers, complex thermal hardware, and contamination-prone workflows. Here, we present an autonomous robotic PCR platform that redefines thermocycling as a motion-controlled process rather than a temperature-controlled device. The system employs a programmable robotic liquid handler to execute PCR entirely within sealed pipette tips, repeatedly immersing and withdrawing reaction volumes in a single temperature-stabilized oil bath to realize denaturation, annealing, and extension steps through precise spatiotemporal control. This architecture eliminates conventional thermocyclers and enables fully enclosed reactions with complete sample recovery. We demonstrate that the robotic system achieves amplification efficiency and sequencing fidelity comparable to high-performance commercial thermocyclers when applied to DNA-encoded datasets. Beyond performance parity, the platform minimizes consumables consumption, suppresses cross-contamination through physical isolation, and supports parallelization through robotic scheduling rather than hardware duplication. Structural contamination control is demonstrated through sealed-tip confinement, validated by no-template control experiments under deliberate challenge conditions. By abstracting PCR thermocycling into a robotically orchestrated manipulation task, this work establishes a generalizable framework for automated biochemical processing and positions robotic control as a central design axis for scalable, low-cost molecular workflows.

# Introduction

The exponential growth of digital data has created an urgent need for sustainable, long-term storage solutions(*1*). DNA has emerged as a compelling alternative, offering theoretical storage densities exceeding 1 exabyte per cubic millimeter and stability spanning millennia under appropriate conditions (*2*, *3*). Unlike conventional storage media that degrade within decades, DNA's inherent molecular stability and information density make it uniquely suited for archival applications requiring century-scale data preservation (*4*, *5*).

Despite significant advances in DNA data storage technology, the transition from laboratory demonstrations to practical archival systems remains constrained by workflow complexity. The complete pipeline—encompassing synthesis, storage, retrieval, amplification, library preparation, and sequencing—requires multiple specialized instruments and protocols (*6*–*8*). Widespread adoption by archival institutions demands integration of these steps into autonomous, end-to-end platforms requiring minimal human intervention.

Amplification is a critical step in this workflow, essential for both data retrieval and re-amplification of archived samples (*9*, *10*). However, unlike diagnostic or research applications where speed is prioritized, DNA data storage workflows demand reliability, contamination control, and long-term operational sustainability over throughput. These requirements create an opportunity to fundamentally rethink amplification strategies for autonomous operation and robotic integration.

To develop an automation-compatible amplification approach, we revisited PCR's evolution to identify design principles suited for robotic implementation. The original PCR protocol developed by Kary Mullis employed manual sample transfers between three temperature-controlled water baths—a simple yet effective technique that achieved rapid thermal transitions through direct sample-to-bath contact (*11*, *12*). Subsequent automated thermocyclers transformed PCR accessibility, with current high-performance instruments achieving heating rates approaching 5 °C/s through sophisticated thermal control systems (*13*, *14*). However, these devices function as standalone units that resist integration into robotic workflows, requiring separate handling steps and additional equipment that compound system complexity.

Microfluidic platforms emerged as an alternative, leveraging enhanced heat transfer from increased surface-to-volume ratios to enable integrated sample processing (*15*–*17*). Digital PCR (dPCR) extended this paradigm by partitioning reactions for absolute quantification with improved sensitivity (*18*, *19*). Despite these advances, microfluidic approaches are difficult to integrate with DNA data storage workflows, primarily because complete sample retrieval from microfluidic devices—essential for downstream sequencing—is challenging (*20*). Furthermore, their limited flexibility impedes adaptation to evolving protocols, as modifications typically necessitate new chip designs rather than simple software updates (*21*, *22*).

Current automation-compatible solutions partially address these bottlenecks but introduce additional instrumentation. Commercial robotic liquid-handling platforms, such as Opentrons, offer programmable automation yet require multiple liquid transfers between pipette tips, PCR tubes, or plates, followed by amplification in separate thermocyclers (*23*, *24*). This multi-step workflow demands both robotic handlers and dedicated thermocyclers, while introducing contamination risks and mechanical complexity through repeated transfers and lid manipulations (*25*).

Here, we introduce a disruptive approach that revisits the original water bath-based PCR method and integrates it with modern robotic liquid-handling technology, eliminating equipment redundancy. Our platform, termed PCRobot, employs a robotic unit with an integrated pipettor to perform amplification via controlled immersion and withdrawal of sample-loaded tips in a single heated oil bath. This configuration achieves heating rates comparable to advanced thermocyclers (~5 °C/s) while exploiting passive air convection for cooling.

Amplification occurs entirely within sealed pipette tips, with temperature plateaus controlled by modulating immersion duration rather than complex thermal hardware. This architecture eliminates contamination-prone transfers, enables quantitative sample recovery, and reduces mechanical complexity, equipment redundancy, and consumable waste. The efficient heat transfer from direct oil-to-sample contact, combined with compatibility with standard robotic infrastructure, yields a scalable, automation-ready amplification platform. Structural contamination control is further validated experimentally through no-template control experiments performed under deliberate challenge conditions.

Using DNA-encoded datasets, we demonstrate that PCRobot achieves amplification performance comparable to conventional automated workflows while reducing contamination risk and minimising instrumentation. By embedding thermal cycling directly within robotic operations, PCRobot represents a paradigm shift toward fully integrated, autonomous DNA data storage systems optimized for archival deployment.

# Results

## PCRobot: Robotic liquid handling integrated with thermal cycling.

We developed an automated amplification platform designed for integration into end-to-end DNA data storage workflows (Fig. **1a**). The system combines a robotic arm-mounted pipettor for liquid handling (mixing, dispensing) with in-tip thermal cycling. Reagents (primers, template DNA, PCR master mix) are mixed through automated pipetting, but rather than transferring the mixture to PCR tubes for amplification in a separate thermocycler, the reaction mixture remains in the pipette tip, which serves as a high surface-to-volume ratio reaction vessel. Temperature cycling is achieved by immersing the tip in a heated oil bath (heating) and withdrawing it into ambient air (cooling).

Our approach revisits early PCR protocols that employed manual tube transfers between temperature-controlled water baths for denaturation, annealing, and extension. Although we initially tested a three-bath configuration (Supplementary Material and Fig. **S1**), we adopted a simpler single-bath design maintained above the highest target temperature. Temperature plateaus are achieved using pulse-width modulation: the tip undergoes repetitive immersion and withdrawal cycles to maintain target temperatures, producing characteristic oscillations around setpoints (Fig. **1b,c**). Both duty cycle (immersion/withdrawal time ratio) and immersion depth can be modulated to control temperature. Temperature monitoring during this optimisation process employed a custom thermocouple assembly (detailed below in "Mock pipette with thermocouple for in-situ characterization").

To minimize footprint while maintaining autonomy, the MECA500 robot and Seyonic single-channel pipettor—connected via a custom holder—are assembled on a 50 × 23 cm breadboard integrating holders for consumables: tip rack, tubes (10×), custom sealing caps (24×; see "*Sealing to retain liquids during thermal cycling*"), and waste receptacle (Fig. **1** and Fig. **S2**).

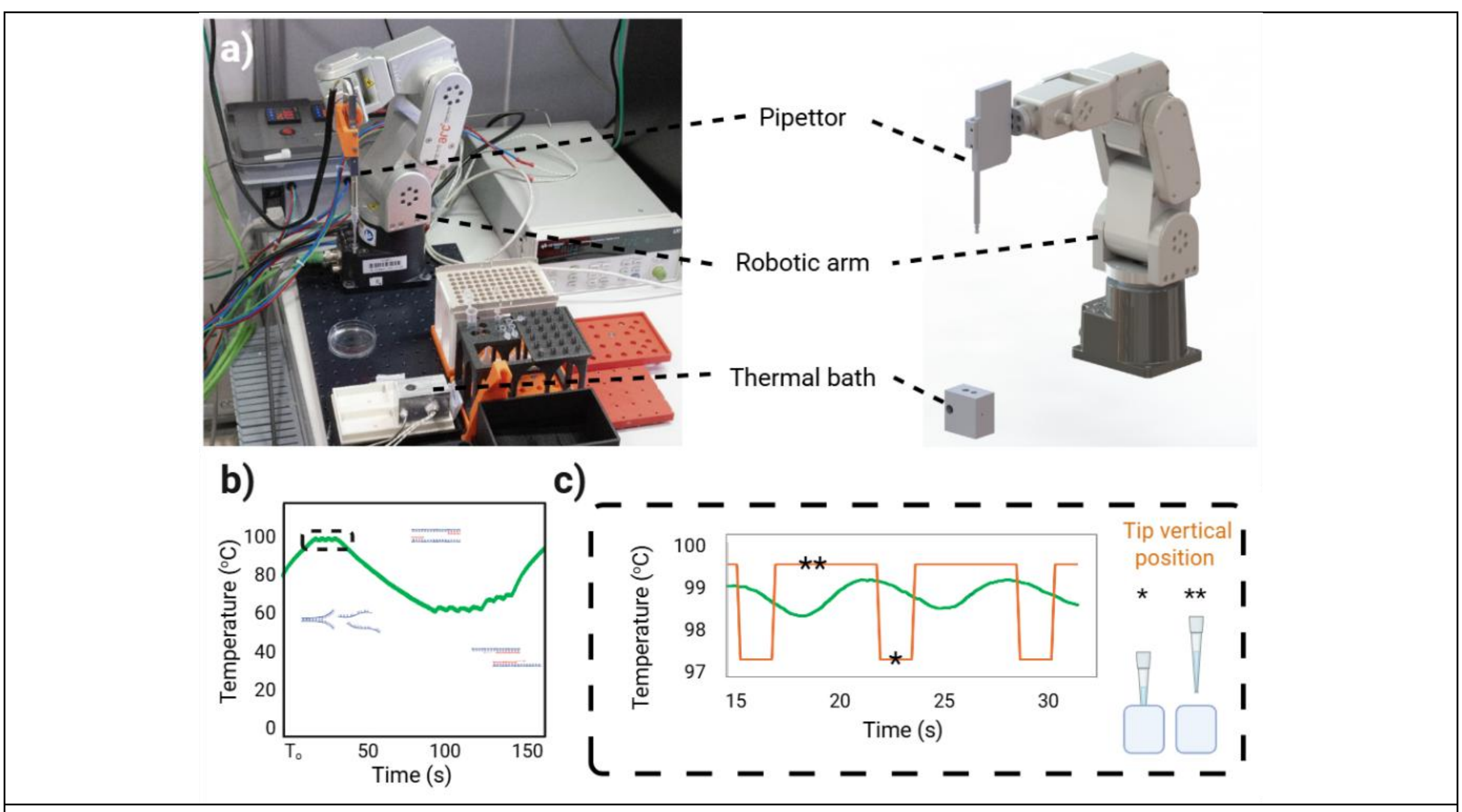

*__Figure 1__. PCRobot platform and thermal cycling principle. (a) Overview of the system comprising the MECA500 robotic arm, Seyonic single-channel pipettor, and temperature-controlled oil bath, assembled on a 30 × 60 cm breadboard. (b) Temperature profile during a single PCR cycle, showing denaturation, annealing, and extension plateaus achieved through controlled tip immersion and withdrawal. (c) Relationship between tip vertical position (qualitative) and measured temperature, illustrating the temporal offset arising from thermal inertia and acquisition latency.*

A key advantage of this architecture is the reduction in workflow complexity relative to conventional thermocycler-based amplification. Table **1** details and compares the process steps for both workflows, from PCR mix preparation (step 0) through to purified product (step 6). By confining the reaction within a single sealed pipette tip throughout thermal cycling, PCRobot eliminates the sample transfer, vessel change, and instrument loading steps that are inherent to thermocycler-based workflows.

**Table 1**. Workflow comparison between conventional thermocycler (manual) and PCRobot for a single PCR amplification reaction, from mix preparation to purified product.

| # | Step | Thermocycler (manual) | PCRobot |
|---|---|---|---|
| 0 | *PCR mix preparation* | Manual (aspiration, mixing, tip changes, open tube) | *Automated (aspiration, mixing, tip changes, open tube)* |
| 1 | Sample transfer to reaction vessel | Manual transfer to PCR tube; open tube | No transfer; reaction remains in tip |
| 2 | Reaction vessel sealing | Manual tube-cap closure | Automated cap placement by robot |
| 3 | Transfer to thermocycler | Manual placement into thermocycler block; lid opening | No transfer; tip immersed directly in bath |
| 4 | Thermal cycling | Automated (thermocycler) | Automated (robotic immersion/withdrawal) |
| 5 | Reaction vessel opening post-cycling | *Manual lid removal; open tube* | Automated cap removal by robot after cooling |
| 6 | Sample recovery from reaction vessel | *Manual aspiration from PCR tube; tip change* | Direct dispensing from tip; no transfer needed; tip change |

## Mock pipette with thermocouple for in-situ characterisation.

Miniaturized Type-K thermocouple junctions were fabricated through controlled micro-fusion of fine-gauge Chromel–Alumel wires, yielding low-thermal-mass sensors for localized temperature measurements (Fig. **S3** and Methods). These were integrated into a mock mandrel designed to replicate Seyonic single-channel pipettor dimensions. The same thermocouple junctions characterized laboratory thermocyclers for comparison and to establish target temperature tolerances (Fig. **S4**).

Extensive design iterations ensured reliable tip insertion and air-tight sealing. The final design employs a shortened mandrel with friction fit using two O-rings. O-rings compensate for relatively rough 3D-printed surfaces, ensuring proper tip sealing. The shortened design reduces displaced air volume during tip insertion, lowering initial internal pressure and preventing tip detachment during thermal cycling.

## Sealing tips retains liquid during thermal cycling.

Initial PCRobot evaluation revealed that regardless of tip type or volume, liquid retained during pipetting at ambient temperature was expelled above 40 °C (Fig. **2a,b**). We systematically investigated this phenomenon. Our initial hypothesis—thermal expansion of fluids (solution and trapped air)—could not fully explain liquid loss (Supplementary Material and Fig. **S5**). The most likely cause is combined thermal expansion with temperature-dependent changes in surface tension (*26*–*28*) and viscosity (*29*), both of which decrease with temperature and negatively impact liquid retention.

We addressed this by sealing tips with custom elastomeric caps during thermal cycling. The caps were designed with an undersized aperture to ensure a tight friction fit and airtight seal around the tip opening, and a conical upper surface to guide tip insertion under minor misalignment. Caps were fabricated by injection moulding using Santoprene thermoplastic vulcanizate, selected for its mechanical properties and its compatibility with high-volume production (Fig. **S6**). During operation, the robot automatically retrieves caps before amplification and removes them 30 seconds after the final cycle, ensuring temperature drops below 40 °C to prevent heat-related liquid loss.

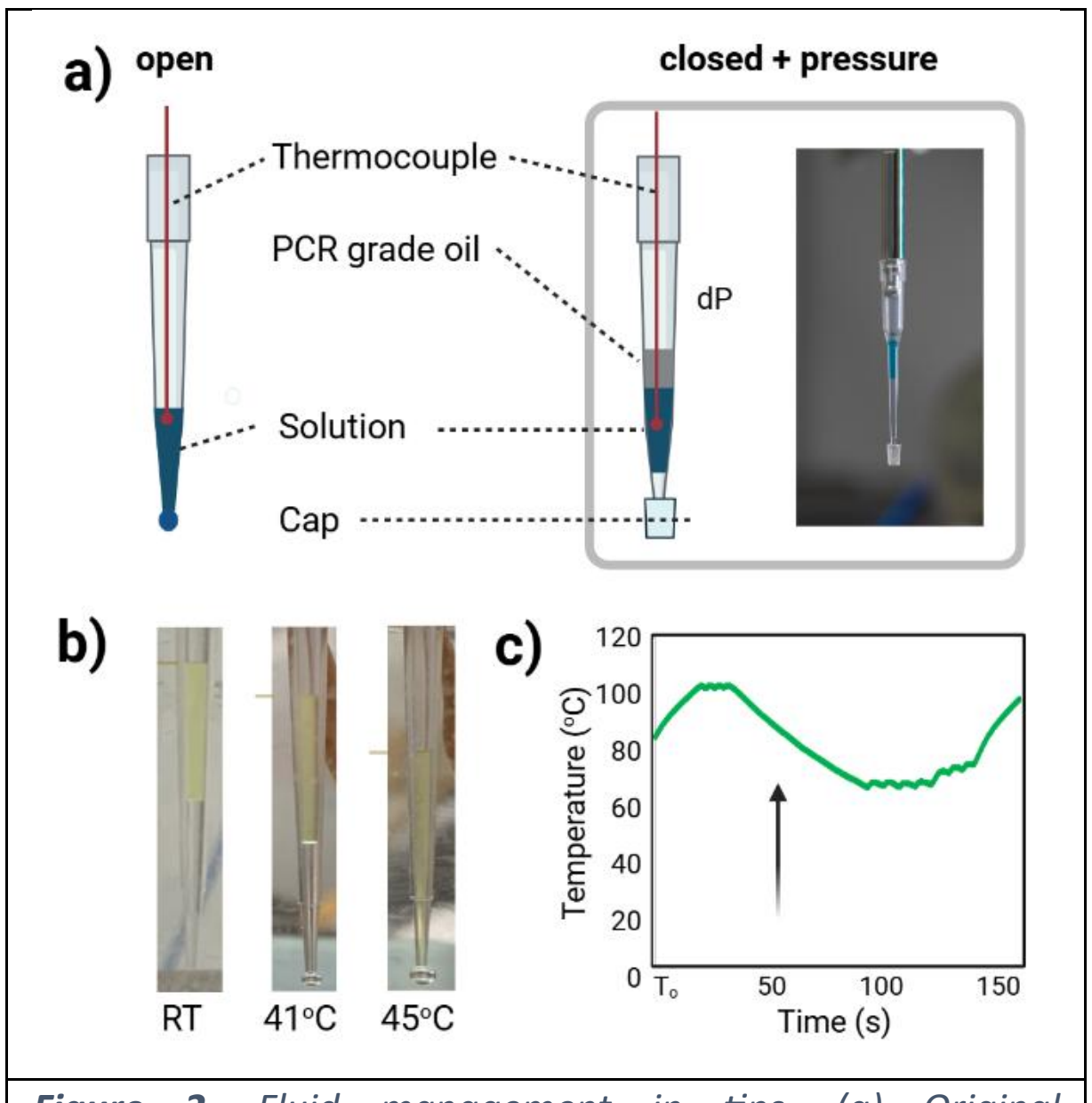

***Figure 2.*** *Fluid management in tips. (a) Original configuration (left) caused liquid loss above 40 °C (b). Adding caps prevented liquid loss but generated air bubbles during cooling (c). Adding PCR-grade oil and applying positive pressure resolved both issues, preventing evaporation.*

This approach achieved complete liquid retention over 25 PCR cycles. However, air bubbles formed systematically during transitions from denaturation (96 °C) to annealing (74 °C), indicating that rapid thermal expansion and contraction generate bubbles even with degassed solutions (Fig. **2c**). When liquid was fully submerged in the bath, bubble formation interfered with thermal measurements and process optimization. Additionally, liquid evaporation occurred inside tips. We addressed both issues simultaneously by overlaying PCR-grade oil on the reaction mixture and applying positive pressure (1200 mbar) during operation. This optimized configuration maintained stable conditions over 25 cycles.

## PCRobot enables structural contamination control.

Amplicon carryover and reagent-derived DNA contamination are well-documented sources of error in PCR and NGS workflows. Automated systems employing amplification and detection in a closed system have substantially reduced the possibility of false-positive results due to amplification product carryover. In multi-step PCR library preparation workflows, cross-contamination from amplicons can arise from both the template and the PCR reagents themselves (*30*, *31*).

PCRobot addresses this risk structurally rather than procedurally. As summarised in Table **2**, a conventional manual thermocycler workflow requires at least three open-tube events, two tip or tube changes, and three liquid transfers — each a potential contamination entry point. PCRobot reduces these to one open-tube event, one tip change, and two liquid transfers, with the reaction confined entirely within a sealed pipette tip throughout thermal cycling (see Table **1**).

This containment is further reinforced by the fluid management strategy. The overlay of PCR-grade oil combined with positive pressure (1200 mbar) creates a physical barrier preventing vapour or aerosol migration into the pipettor body, and ensures net airflow is directed outward from the tip during operation. Eliminating the transfer of pre-amplified sample between vessels removes a critical contamination risk in sequential amplification workflows(*32*).

To validate contamination control under challenging conditions, we performed a no-template control (NTC) experiment in which 1 ng of DNA was deliberately introduced into the oil bath to simulate carryover from a prior amplification. No amplification was detected, as confirmed by electrophoregrams (Fig. **S7**), demonstrating that the oil bath does not act as a contamination reservoir between sequential runs.

**Table 2**. Contamination risk metrics extracted from the amplification workflow (steps 1–6, Table **1**), comparing a conventional manual thermocycler setup with PCRobot. Open-tube steps, tip and tube changes, and liquid transfers represent discrete contamination entry points; lower values indicate reduced contamination risk.

| Metric | Thermocycler (manual) | PCRobot |
|---|---|---|
| Total open-tube steps | 3 | 1 |
| Total tip/tube changes | 2 | 1 |
| Total liquid transfers | 3 | 2 |
| Instruments required | Thermocycler + pipette | PCRobot only |

## Thermal efficiency and process time optimisation.

Having established the platform's sample integrity and contamination control properties, we next characterise its thermal performance and amplification efficiency. To this end, biologically meaningful temperature tolerances for PCRobot were established by directly measuring the thermal fluctuations occurring within a PCR tube during conventional thermocycling using our custom miniaturised Type-K thermocouple assembly (Fig. **S4**). These measurements revealed that commercial thermocyclers themselves operate with inherent temperature variability across all three plateaus. The ±2 °C tolerance adopted for PCRobot was therefore set to match the performance envelope of the commercial instrument used as benchmark, ensuring that both systems are compared on equivalent thermal terms and that the tolerance criterion reflects an instrument-grounded, biologically validated threshold.

Figure **3** compares time distributions around target temperature plateaus (red lines) for PCRobot and a conventional thermocycler. Both systems show comparable performance, although PCRobot demonstrates higher accuracy (particularly for annealing and denaturation) but slightly reduced precision. Broader temperature dispersion during annealing and expansion are due to a faster temperature rise at lower temperature, creating a higher amplitude temperature variation within a step. This effect was mitigated by reducing the immersion depth during those two steps.

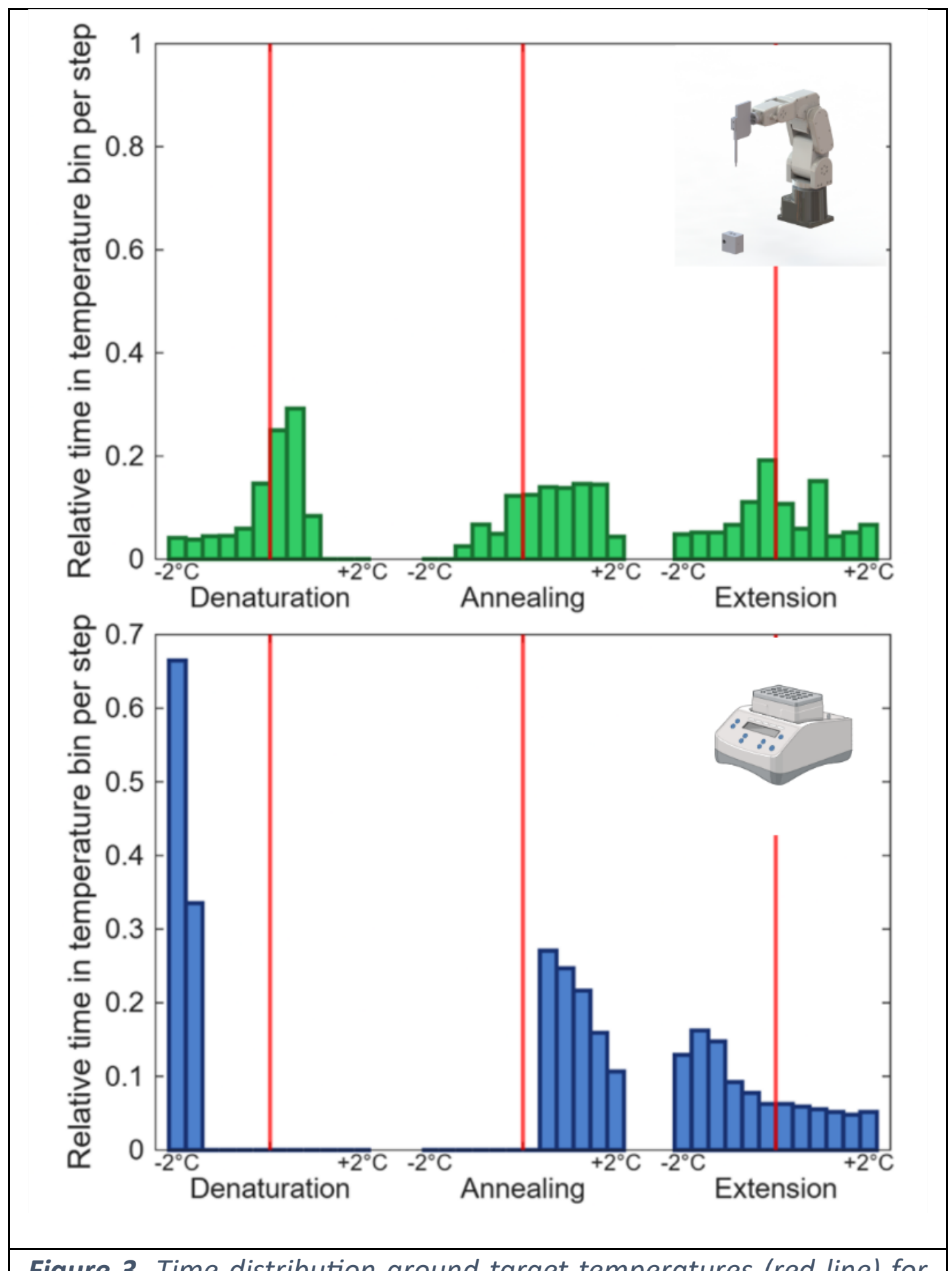

***Figure 3.*** *Time distribution around target temperatures (red line) for PCRobot (a) and thermocycler (b) over 5 cycles. Counts were normalized to a total of 1 for each step. Each histogram spans ±2 °C around the setpoint (red vertical line), divided into 12 bins.*

While PCRobot achieves heating performance comparable to commercial thermocyclers through direct oil bath contact, it is outperformed during cooling (Fig. **4**). For 25 cycles, the thermocycler requires approximately 30 minutes versus over 60 minutes for PCRobot. This difference arises because PCRobot relies on passive air convection, whereas thermocyclers employ active cooling. Despite the longer processing time, PCRobot's energy consumption (0.054 kWh), as measured by a power meter, is only marginally higher than that of the thermocycler (0.043 kWh). This comparison, however, is not performed on equivalent terms. PCRobot is a fully autonomous system that integrates a six-axis robotic arm, a precision pipettor, and a thermostated bath — no human intervention is required from reagent aspiration to amplification completion. In contrast, the thermocycler measurement reflects only the amplification device itself, and excludes the robotic liquid handler that would be required to load samples in an equivalent automated workflow. Commercial robotic liquid handlers such as the Opentrons OT-2, which are commonly paired with thermocyclers in automated pipelines, consume on the order of 60–100 W during operation, which over a 30-minute loading and preparation cycle would add approximately 0.03–0.05 kWh to the thermocycler-based workflow. When accounting for this, the total energy footprint of a thermocycler-based automated workflow becomes comparable to or exceeds that of PCRobot.

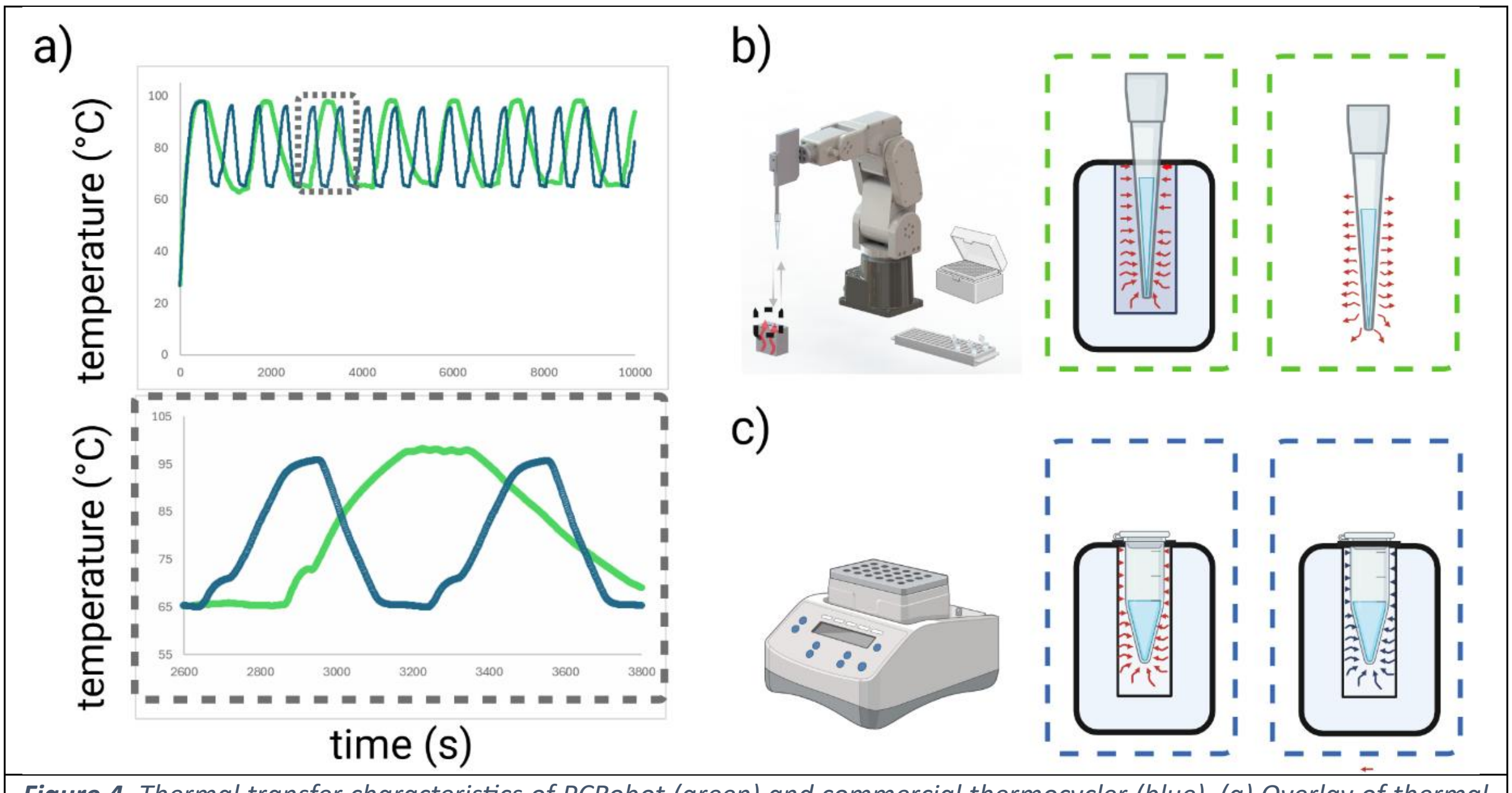

***Figure 4.*** *Thermal transfer characteristics of PCRobot (green) and commercial thermocycler (blue). (a) Overlay of thermal cycles for each system (see Fig.* ***S4*** *for details). Lower panel shows detail of thermal cycles, illustrating slower PCRobot cooling via air convection (b) compared to active cooling in the thermocycler (c).*

Although timing is not critical for archival DNA data storage, other applications may benefit from reduced cycle times. Using a three-bath configuration with a faster XYZ robotic platform, we achieved 20 cycles in 22 minutes (Supplementary Fig. **S1c**). Further reduction could be achieved by adding a fourth temperature-controlled bath below the lowest target temperature. First order impulse response measurements, $T(t) = T_1 + (T_0 - T_1)(e^{-t/\tau})$, quantified cooling in ambient air (24°C) versus a regulated bath (51°C). The time constant τ decreased respectively from 43s to 6s when cooling from 96°C to 54°C (see SM; Figure **S5** and Table **S1** for details). While this represents substantial time savings, it would increase system complexity, requiring additional baths, thermal controllers, and robotic steps.

An alternative approach to improve thermal transfer would be reducing tip wall thickness. The material and geometric parameters of the pipette tip represent critical factors influencing heat transfer efficiency, and wall thickness in particular has a dominant effect on thermal time constants under immersion conditions. We evaluated potential benefits by measuring the instantaneous inverse thermal time constant (1/τ) for standard pipette tips versus PCR tubes immersed in a heated bath (see Methods). PCR tube walls (0.3 mm) are half the thickness of standard tips (0.6 mm), and Fig. **5** shows that PCR tubes exhibited consistently higher 1/τ values across the full temperature range, demonstrating substantially faster thermal equilibration. Both vessels displayed non-monotonic temperature dependence: 1/τ increased at lower temperatures, peaked around 40–45 °C, and decreased at higher temperatures, reflecting temperature-dependent water properties (viscosity, thermal conductivity) and potential boundary-layer effects at the vessel–fluid interface. Importantly, the relative offset between curves remained consistent across the temperature range, confirming that wall thickness dominates thermal time constants under these conditions and that thinner-walled tips would directly improve thermal cycling performance.

Tip selection for PCRobot was therefore performed with thermal transfer efficiency as a primary criterion, alongside two hardware constraints: compatibility with the Seyonic single-channel pipettor, and sufficient internal volume to accommodate the 25 µL reaction. Among the available tip designs satisfying both constraints, the most slender form factor readily available was selected to maximise the surface-to-volume ratio and minimise wall thickness. Further improvement could be achieved by developing custom thin-walled tips.

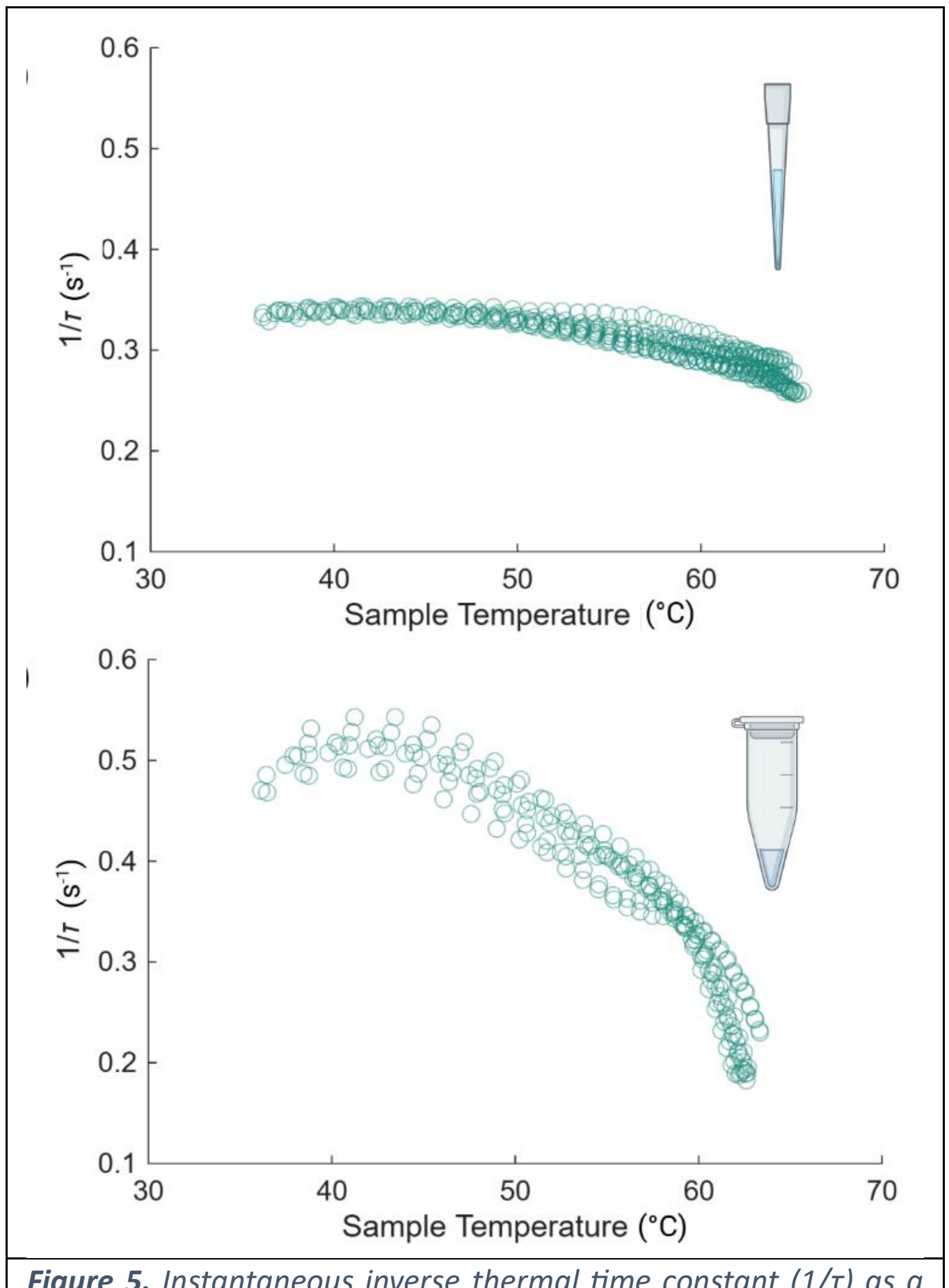

***Figure 5.*** *Instantaneous inverse thermal time constant (1/τ) as a function of internal liquid temperature for pipette tip (a) and PCR tube (b). Both containers show similar non-monotonic temperature dependence. PCR tubes exhibit consistently higher 1/τ values, attributed to thinner walls.*

## Amplification efficiency comparable to commercial thermocycler.

We compared the amplification results of the PCRobot to a commercial thermocycler, under identical conditions: 25 µL reaction, 25 cycles. Post-purification DNA concentrations differed significantly between PCRobot (21.6 ng/mL) and the conventional thermocycler (13.1 ng/mL; two-tailed t-test: t(4) = −3.07, p = 0.037). However, amplification efficiency ($E$), defined in $C_m = C_0(1+E)^m$, with $C_0$, the initial concentration, $C_m$, the concentration after $n$ cycles, provides a more appropriate comparison metric. PCRobot demonstrated significantly higher amplification efficiency (0.686 ± 0.011, n = 3) compared to the thermocycler (0.662 ± 0.003, n = 3; t(4) = −3.72, p = 0.020). This difference, while statistically significant, is marginal in practical terms as shown in Figure **6**a. Electropherograms confirmed that expected PCR products (amplicons) were equivalent for both systems (Fig. **S8**).

To assess the generalisability of PCRobot across oligonucleotide templates of varying length and sequence composition, we performed two independent comparative runs using a test DNA pool comprising 500 nt oligonucleotides. Both systems achieved higher overall amplification efficiencies than observed for the 243 nt pool, consistent with the longer template. Across both runs, PCRobot (0.756 ± 0.014, n = 3) and the thermocycler (0.775 ± 0.017, n = 4) showed comparable performance, with no statistically significant difference between platforms (two-tailed t-test: t(5) = −1.58, p = 0.175) as shown in Fig **6b**. The mean difference of 0.019 falls well within the range of experimental variability, and the consistency of results across two independent replicate sets further confirms the robustness of the PCRobot approach for longer oligonucleotide templates. The expected PCR products (amplicons) were equivalent for both systems as confirmed by electropherograms (Fig. **S7**). Additional amplicon integrity for a 304 nt fragment was confirmed using the MiniOne® Electrophoresis Reagent Starter Kit (M3200, The MiniOne), which positively evaluated PCRobot performance (Fig. **S1c**).

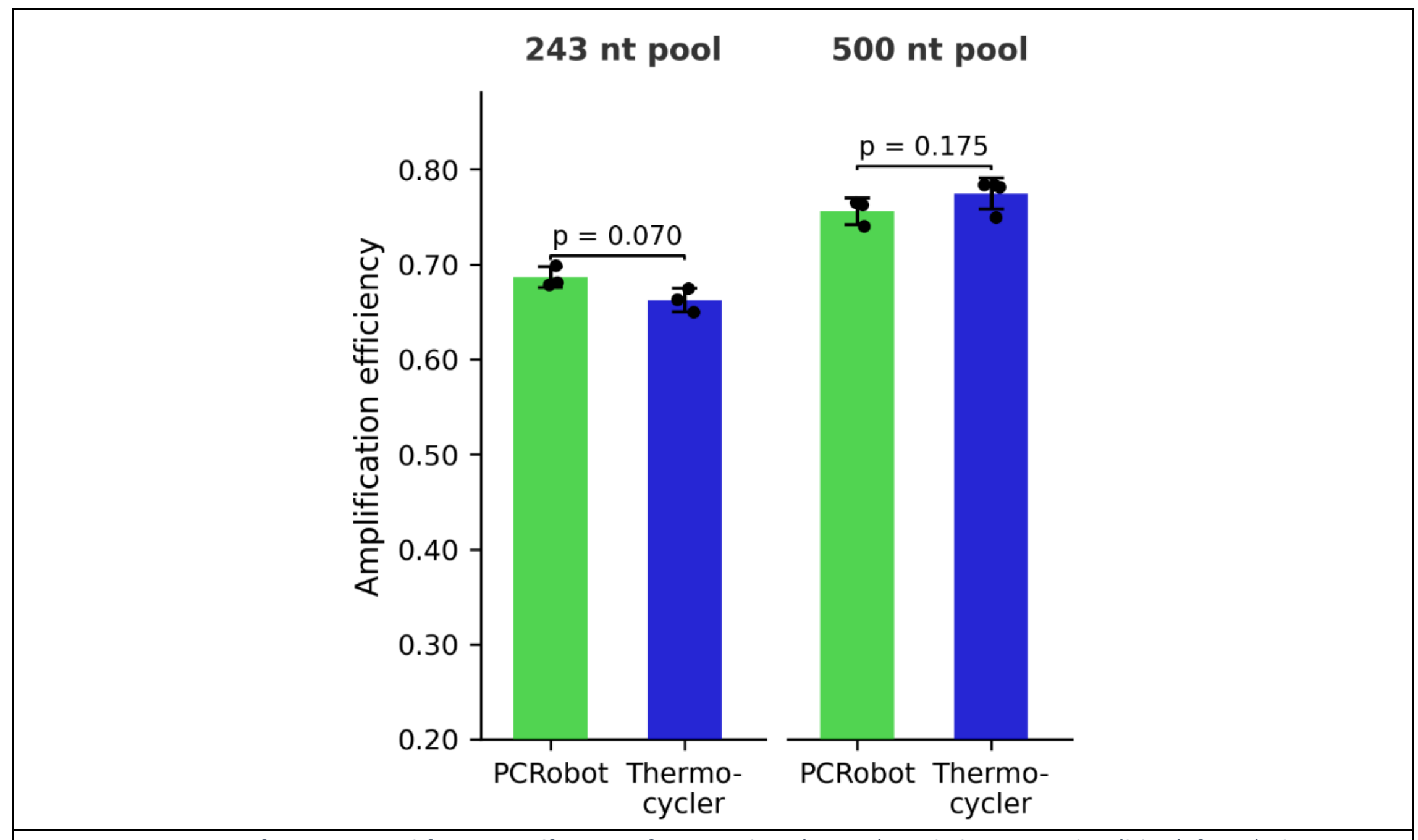

***Figure 6**. Post-purification amplification efficiency for PCRobot (green) and thermocycler (blue) for a) the 243 nt oligonucleotide pool (mean efficiencies 68.6% and 66.1% respectively; two-tailed t-test: t(4) = −3.72, p = 0.020) and b) the 500 nt oligonucleotide pool (mean efficiencies 75.6% and 77.5% respectively; two-tailed t-test: t(5) = −1.58, p = 0.175). Individual data points are shown.*

## Validation for DNA data archival workflows

We validated the automation-compatible amplification system within a DNA data storage workflow using Oxford Nanopore Technologies (ONT) sequencing. The validation dataset consisted of the DNAMIC logo (*33*) encoded in SVG format, encapsulated in an Open Archival Information System (OAIS, ISO 14721)–compliant OLOS package (*34*). The file (8.1 kB) was encoded using a custom codec (Supplementary Materials and Fig. **S9**), incorporating 30% Reed–Solomon error-correction redundancy, yielding 352 oligonucleotides (243 nt each). Each oligonucleotide contained 18-nt forward and reverse primers flanking a 207-nt data payload, designed according to thermodynamic and hybridization constraints to ensure amplification efficiency and primer orthogonality. Figure **7a** illustrates the complete DNA data storage workflow applied to the DNAMIC logo, encompassing encoding, synthesis, storage, amplification, library preparation, sequencing, and decoding. The amplification step of the encoded oligonucleotide pool was performed using both a standard thermocycler and PCRobot under identical cycle numbers and reaction compositions. Following amplification, ONT library preparation and sequencing were performed. Sequencing reads were aligned to the reference oligonucleotide library to assess amplification uniformity, yield, and fidelity.

## Nanopore sequencing reveals comparable amplification uniformity between PCRobot and thermocycler.

Next, we compare multiple quality metrics for multiplex amplification and sequencing between platforms. Read quality scores were similarly high, with median values around 20 for both PCRobot and thermocycler (Fig. **7b**). The number of reads passing quality filtering ($qs > 9$) showed no significant difference (Wilcoxon test, $p > 0.05$; Fig. S**10a**), indicating PCRobot does not compromise sequencing data quality. Alignment completeness, measured as the fraction of read length mapped to reference sequences, displayed nearly identical distributions, with most reads achieving complete alignment (Fig. S**10b**). Total reads mapped to payload sequences did not differ significantly between platforms (Wilcoxon test, $p > 0.05$; Fig. **7c**). Coverage uniformity across payloads, quantified using the Gini coefficient, revealed no significant difference in amplification bias (Wilcoxon test, $p > 0.05$; Fig. S**10c**). Lorenz curve analysis illustrated this equivalence, with both devices showing similar deviation from perfect uniformity (Fig. **7d**). Empirical cumulative distribution of read counts per payload demonstrated comparable performance, with largely overlapping distributions indicating similar payload representation (Fig. S**10d**). Collectively, these results demonstrate that PCRobot performs equivalently to conventional thermocyclers for oligonucleotide pool PCR amplification, producing comparable data quality, yield, and target coverage uniformity.

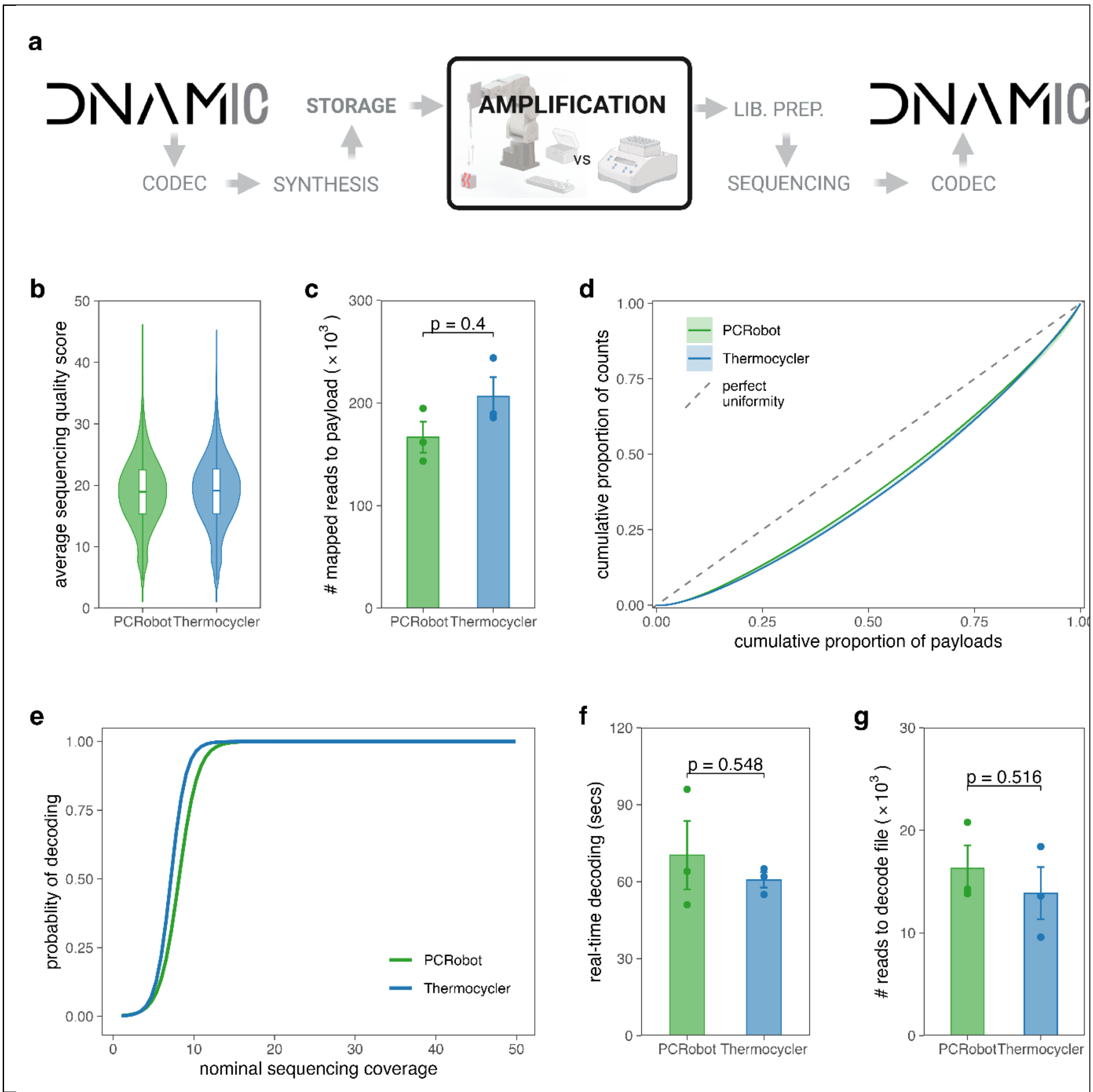

***Figure 7.*** *Performance comparison of PCRobot and Thermocycler for oligopool PCR amplification and sequencing in the context of DNA data storage. (**a**) The workflow encompasses the digital dataset (DNAMIC (33) logo in SVG format), encoding, synthesis and storage. Samples were split for parallel amplification using thermocycler versus PCRobot, followed by library preparation, sequencing, decoding, and readout. (**b**) Distribution of average sequencing quality scores for reads generated from PCRobot and Thermocycler PCR amplification. Violin plots show the density distribution overlaid with boxplots (white boxes, median and interquartile range). (**c**) Total number of reads mapped to payload sequences (allowing max four mismatches). Bars represent mean ± SEM across samples (n = individual samples shown as points). Statistical significance assessed by Wilcoxon test. (**d**) Lorenz curves showing the cumulative distribution of read counts across payloads for each device. Lines represent mean counts per payload with shaded ribbons indicating ± 1 SD. Dashed grey line represents perfect uniformity. Greater deviation from the diagonal indicates less uniform coverage. (**e**) Probability of successful decoding as a function of nominal sequencing coverage for PCRobot and Thermocycler samples. Lines represent generalized linear model fits (quasibinomial) showing the relationship between raw sequencing coverage and decoding success rate for each device. (**f**) Real-time decoding latency measured as the time interval (in seconds) between sequencing initiation and successful decoding. Bars represent mean ± SEM across samples (n = individual samples shown as points). Statistical significance assessed by two-sided t-test. (**g**) Number of sequencing reads required to successfully decode the encoded message in real time. Bars represent mean ± SEM across samples (n = individual samples shown as points). Statistical significance assessed by two-sided t-test. All data represent replicates with PCRobot (blue) and Thermocycler (green) conditions.*

### File decoding performance is comparable between PCRobot and thermocycler.

To assess whether PCR platform choice affects data decoding, we evaluated real-time decoding performance for samples amplified using PCRobot versus thermocycler. Probability of successful decoding as a function of nominal sequencing coverage showed similar trends for both devices, with success increasing with coverage depth (Fig. **7e**). Generalized linear modeling revealed that decoding success saturated (10/10 successful decodings from random sampling) at approximately 10× coverage for thermocycler and approximately 12× coverage for PCRobot, indicating slightly higher coverage requirements for consistent decoding with the automated platform. Real-time decoding latency, measured as the time interval between sequencing start and successful decoding, showed no significant difference between platforms (t-test, $p > 0.05$; Fig. **7f**), although PCRobot exhibited a slight trend toward longer decoding times. Similarly, the number of sequencing reads required for successful decoding did not differ significantly between PCRobot and thermocycler (t-test, $p > 0.05$; Fig. **7g**), despite a modest trend toward higher read requirements for PCRobot samples.

# Discussions

Robotics and automation are transforming molecular workflows in life sciences, not merely by increasing throughput but by enabling fundamental reconsideration of how core biochemical processes are implemented. Rather than replicating established laboratory practices, automation provides an opportunity to redesign protocols for robustness, integration, and sustainability. Here, we applied this paradigm to PCR amplification by revisiting water bath–based thermal cycling and integrating it directly within a robotic platform.

PCRobot implements thermal cycling as a robotic operation through controlled immersion and withdrawal of sealed pipette tips in a single temperature-controlled bath. Evaluation using DNA-encoded datasets demonstrates that this architecture preserves the essential molecular properties required for DNA data storage—reliable amplification, uniform representation of encoded sequences, and sequencing fidelity—at levels equivalent to conventional thermocycler workflows.

Importantly, PCRobot transforms conventional PCR performance metrics into intrinsic system properties. Contamination control is achieved structurally through sealed-tip amplification and elimination of inter-instrument transfers, rather than procedural safeguards. This structural containment was experimentally validated under deliberate challenge conditions, with no amplification detected in no-template controls exposed to exogenous DNA in the oil bath, confirming that the platform architecture itself prevents contamination rather than relying on operator discipline. Automation compatibility emerges from embedding thermal cycling within the same robotic infrastructure used for liquid handling, eliminating dedicated thermocyclers and associated mechanical interfaces. Energy efficiency results from simplified thermal management—a single heated bath with passive cooling—rather than actively controlled thermal blocks, reducing hardware complexity.

Table **3** consolidates the system-level performance of PCRobot against a conventional thermocycler across the key operational dimensions evaluated in this work and summarises the structural advantages that emerge from embedding thermal cycling within a robotic platform.

**Table 3**. System-level performance comparison across key operational metrics.

| Parametre | Thermocycler (manual) | PCRobot |
|---|---|---|
| Amplification efficiency (243 nt) | 66.3 ± 1.3 % | 68.6 ± 1.1 % |
| Amplification efficiency (500 nt) | 77.5 ± 1.7 % | 75.6 ± 1.4 % |
| Sequencing (read quality score) | 20 | 20 |
| Processing time (25 cycles) | ~30 min. | ~60 min. |
| Energy consumption | 0.043 kWh † | 0.054 kWh |
| Cross-contamination risk | Protocol dependent | Structurally minimised |
| Automation integration | Requires separate robotic unit | Native integration |

† *Energy consumption for thermocycler reflects amplification device only, excluding robotic sample loading*

More broadly, this work demonstrates how reframing molecular biology protocols as components of integrated automated systems can yield architectures that are functionally equivalent at the biochemical level while superior at the system level. This approach may inform redesign of other core molecular operations as life sciences transition toward autonomous, distributed, and archival-scale technological platforms.

# Methods

## Smart Pipettor:

**Pipettor**: We used a Single Channel Pipettor (PCNC-0061-00) from Seyonic SA with a System Controller Dual Pressure/Vacuum Source (Model PCNC-0071-00) for ultra-precise pipetting. The pipetting system is operated remotely via TCP/IP. For the **robotic unit**, we used the Meca500, a compact six-axis industrial robotic arm from Mecademic. It has a reach of 330 mm and a payload capacity of 0.5 kg. The unit was fixed to a 500x230x10 mm breadboard that also accommodated 3D printed parts for supplies and waste. The robot was connected to the pipettor using a custom-made 3D printed holder. **3D printing** was performed using a Prusa mk3s FDM (Fused Deposition Modeling) printers and 1.75 mm PLA filaments from FormFutura. **Control software.** We use the Synapxis integrated development environment (IDE) and control interface, which supports both TCP/IP and serial communication, to control the robot and pipettor. A dedicated script (available in Annex) was developed for the occasion.

The **mock pipette** was designed to be reliably printed on an MSLA 3D printer. To maximize geometric accuracy along the radial axes, the mock pipette parts are printed vertically. To limit the risk of a part unsticking from the print plate, the mock pipette was split into four parts, the mandrel, two “half arms” and one fixation. The parts are connected via a crescent shaped connector, to limit degrees of freedom during construction. The mandrel features a through-hole for the passage of the thermocouple. When connected to the half-arm, the crescent shaped connector leaves a gap between the two parts, allowing the passage of the thermocouple.

**Thermocouples fabrication.** Miniature thermocouple junctions were produced using fine-gauge, commercially available Type-K thermocouple wire (Chromel–Alumel; conductor cross-section 0.05 $mm^2$). The two conductors were gently twisted at the distal end to ensure mechanical alignment prior to junction formation. Electrical energy was delivered via a low-voltage laboratory power supply equipped with an output-side capacitive buffer (five electrolytic capacitors, nominal total capacitance ≈ 1 mF). The positive terminal was contacted through a graphite element wrapped in copper tape, while the negative terminal interfaced with the thermocouple leads through screw-type contacts.

Junction formation occurred when the twisted wire pair was momentarily brought into contact with the graphite element. The localized resistive heating at the contact point promoted rapid melting and fusion of the alloy pair, forming a stable thermo-electric junction. The resulting junctions were visually inspected and tested for electrical continuity and thermal response.

**Thermal transfer efficiency.** Thermal response dynamics of the pipette tip and PCR tubes were quantified during repeated immersion into a thermostated water bath, heated to 71°C and magnetically stirred to ensure spatially uniform temperature. A mock pipette mounted on the robot performed controlled immersion–withdrawal cycles, dipping each vessel to a fixed depth and dwell time. A miniature thermocouple was inserted into the liquid contained within each vessel to monitor the internal temperature, while a second thermocouple positioned adjacent to the immersion point recorded the bath temperature. Both channels were logged at high temporal resolution ($\Delta T = 0.15$ s).

Temperature traces were pre-processed to exclude artefactual points (e.g., transitions between air and water). The instantaneous derivative of internal temperature, $dT_{\text{sample}}/dt$, was obtained numerically. Assuming a lumped-capacitance model, the instantaneous inverse thermal time constant was computed as

$$\frac{1}{\tau}(t) = \frac{dT_{\text{sample}}/dt}{T_{\text{bath}} - T_{\text{sample}}},$$

which reflects the effective heat-transfer rate between the bath and the liquid inside the vessel. The resulting $1/\tau$ values were plotted against the instantaneous internal temperature for each geometry.

**Elastomeric caps**, designed to fit at the end of the tip and stay in place by friction, were fabricated by injection moulding using an *Arburg Allrounder S170* injection press. The selected material was Santoprene™ thermoplastic vulcanizate (TPV), belonging to the thermoplastic elastomer (TPE) family. Prior to injection, the material was dried at 80 °C for 3 hours to eliminate residual moisture. Injection moulding was performed under the following processing parameters: material temperature of 190 °C, mould temperature regulated at 80 °C, and an injected volume of 0.5 $cm^3$ per cycle. The maximum injection pressure was set at 400 bar, and the cycle time was approximately 10 seconds. The mould was designed to ensure uniform filling and consistent part geometry under these thermal and pressure conditions. After ejection, the moulded caps were allowed to cool to ambient temperature before further handling or characterization. All fabrication runs were performed under identical process conditions to ensure reproducibility.

## DNA data:

### Materials/synthesis

**DNA Encoding.** The DNAMIC logo (SVG format, 8.1 kB) was encoded into DNA using the OLOS package with 30% error-correction redundancy, yielding 352 oligonucleotides (oligos). Each oligo comprised 243 nt, including 18-nt forward and reverse primer binding sites flanking a 207-nt data payload. Primer sequences were generated via SHA-256 hashing and selected based on Gibbs free energy, melting temperature difference (< 2 °C), and a Hamming distance > 10 to minimize cross-hybridization. The oligo pool, along with a test oligo pool (500 nt) with random sequence, was obtained from TWIST

Bioscience and primers from Eurofins, additional primers were synthesised using the Kilobaser 1 synthesiser (Mabeal).

**Data Structure and Encoding Scheme.** Input files were converted into an $n \times m$ matrix, where $n = 34$ (equivalent to 153 nts) and $m$ was determined by dividing the file length by 34. Incomplete columns were filled with dummy data generated by a modulo-256 counter to prevent repeating patterns; if $m$ was odd, one column was added. Each sequence included a 4-byte index for positional identification, allowing encoding of files up to 146 GB.

**Error Correction and DNA Conversion.** Error correction employed Reed–Solomon (RS) codes at two levels: inner codes (8 bytes per column) and outer codes (calculated for each row joining two bytes). Redundancy ratios ($r$) were set between 10–30% based on nanopore sequencing performance. Each symbol represented 16 bits encoded by 9 nucleotides using a GF(41) codon wheel ($256^2 \rightarrow 41^3$ base conversion), ensuring that no base was repeated more than three times.

**Thermal cycling PCR amplification** was carried out using the Q5 High-Fidelity DNA Polymerase kit (M0491S; New England Biolabs), following the manufacturer's protocol. Reactions (25 µL final volume) contained template DNA in the nanogram range, forward and reverse primers (10 µM stocks), and the supplied Q5 buffer and dNTPs, with nuclease-free water added to standardize volume. Reactions prepared with M0491S were assembled on ice, as recommended. Because M0494S provides the same Q5 polymerase in a ready-to-use master mix that does not require cold-setup, it represents a fully compatible formulation and will be adopted for future experiments.

Thermal cycling for the logo-encoding DNA pool (234 nt) consisted of an initial denaturation at 98 °C for approximately 30 s, followed by cycles including brief denaturation (≈10 s at 98 °C), annealing (≈15 s at 65 °C), and short extension (≈5 s at 72 °C), with a final elongation step of ~2 min at 72 °C. These parameters were optimized for the DNAMIC Logo amplicon and are consistent with manufacturer recommendations for short, high-fidelity targets. Amplicons were evaluated by agarose gel electrophoresis to confirm fragment length and specificity. For the 500 nt DNA pool, the annealing is at 58°C for 15 s and the extension at 72 °C for 15 s.

The protocol was carried out using the PCRobot and Bio-Gener (Labgene) thermocycler, additional amplification was performed on the MiniOne® PCR system thermocycler (The MiniOne, 2025), using the MiniOne® Electrophoresis Reagent Starter Kit (M3200, The MiniOne) and the 304 nt fragment.

**Purification with MagSi DNA-mf magnetic silica beads**. All steps were performed at room temperature (20–25 °C) using low-bind plasticware and nuclease-free water. *Bead equilibration*. The MagSi DNA-mf bead suspension (silica-based magnetic particles) was homogenized by gentle inversion. An aliquot of 50 µL bead suspension was transferred to a 1.5 mL tube, mixed with 50 µL nuclease-free water, briefly pipette-mixed, and placed on a magnetic rack until the supernatant cleared (~1–2 min). The supernatant was discarded and the equilibration wash was repeated 3 times. *DNA binding*. To the equilibrated beads, 100 µL of "prepared binding buffer" was added and mixed by pipetting. The prepared binding buffer consisted of 20% (v/v) of a stock binding buffer—4 M guanidinium chloride, 40 mM Tris-HCl, 17.6 mM EDTA, pH 8.0 (adjusted with HCl)—diluted in 80% (v/v) isopropanol. Next, 10 µL of sample was dispensed directly into the liquid phase without dislodging the bead pellet or aspirating beads. The tube was vortexed at high speed for 2 min, then placed on the magnet until clear; the supernatant was discarded. *Wash*. A 100 µL wash was performed by carefully adding wash solution over the pellet to avoid disturbing it; the wash solution consisted of 10 mM Tris-HCl, 1 mM EDTA in 80% (v/v) ethanol, pH 8.0. The tube was returned to the magnet and the supernatant was removed. The ethanol wash was repeated (see note below on repeats). *Drying*. Tubes were left open to air-dry the bead pellet for 15 min to evaporate residual ethanol. *Elution*. 20 µL nuclease-free water was added, the tube was shaken for 2 min to resuspend the beads, then placed on the magnet and the eluate was collected into a fresh tube. DNA was stored at −20 °C.

**Quality control.** Amplified DNA oligo pools were quantified using Qubit dsDNA High Sensitivity reagents (Thermo Fisher Scientific), and amplicon size distributions were assessed on an TapeStation 2200 (Agilent) with D1000 ScreenTape.

**ONT nanopore library preparation and sequencing.** For each amplified and quality-controlled oligo-pool sample (thermal cycler, n = 3; PCRobot, n = 3), 250 fmol of purified amplicon DNA was used for ONT library preparation following the high-performance protocol (*8*). Full protocol is available at protocols.io (dx.doi.org/10.17504/protocols.io.dm6gp973jvzp/v2). Sequencing was performed on a MinION Mk1B (ONT) using an R10.4.1 MinION flow cell. To obtain comparable sequencing depth across all six libraries, each run was stopped after reaching approximately 5.5×10^5 reads per sample, yielding ~1,563× nominal coverage. The flow cell was washed and reused between runs according to the manufacturer's instructions. Raw signal data collected in POD5 format was basecalled with Dorado v1.1.1 using the simplex SUP model (dna_r10.4.1_e8.2_400bps_sup@v5.2.0).

**Bioinformatic sequencing analysis**. The resulting sequences in uBAM files were categorized into "passed" or "failed" based on their average quality scores (q-scores), only reads with q-scores above or equal to the threshold of 9 were included in subsequent downstream processing. The reads in uBAMs were converted to FASTQ files using samtools (v1.23) and trimmed for adapters using cutadapt (v5.1) with the following parameters: cutadapt -e 0.04 -g ACGAGCGTGAGTACTCTG...CACCGACTGGAGCTGGCG --action=retain --revcomp -m 235 -M 255. Trimmed reads were then mapped to the reference payload sequences using minimap2 (v2.30) with the following parameters: minimap2 -ax sr -L --MD -t8 -Y –eqx -k10 -w5 -m10 --secondary=yes. Data analysis and visualization were performed using custom Python and R scripts.

**Real-time decoding.** To leverage ONT nanopore sequencing's real-time read output, a Python 3 application with a node-based architecture was developed for real-time decoding. The application monitors MinKNOW's continuous output for newly written raw-read batches, submits each batch to the decoding pipeline, and upon successful decoding, signals MinKNOW via its API to terminate the run. The software is modular (file monitoring, read quality-control, decoding [see Methods], API client) and processes sequential read batches to minimize latency. Batch size of 350 reads was used in this experiment that approximately corresponds to 1X of nominal coverage of the given oligo pool. Implementation details and source code can be provided upon reasonable request.

**Supplementary Materials** :

Figure S1 to S10
Table S1
Movies S1 to S2

# References

1. D. Reinsel, J. Gantz, J. Rydning, The Digitization of the World - From Edge to Core. IDC White Paper. *IDC White Paper* **20**, 3 (2018).

2. G. M. Church, Y. Gao, S. Kosuri, Next-generation digital information storage in DNA. *Science (1979).* **337**, 1628 (2012).

3. N. Goldman, P. Bertone, S. Chen, C. Dessimoz, E. M. Leproust, B. Sipos, E. Birney, Towards practical, high-capacity, low-maintenance information storage in synthesized DNA. *Nature 2013 494:7435* **494**, 77–80 (2013).

4. R. N. Grass, R. Heckel, M. Puddu, D. Paunescu, W. J. Stark, Robust chemical preservation of digital information on DNA in silica with error-correcting codes. *Angew. Chem. Int. Ed.* **54**, 2552–2555 (2015).

5. L. Organick, S. D. Ang, Y. J. Chen, R. Lopez, S. Yekhanin, K. Makarychev, M. Z. Racz, G. Kamath, P. Gopalan, B. Nguyen, C. N. Takahashi, S. Newman, H. Y. Parker, C. Rashtchian, K. Stewart, G. Gupta, R. Carlson, J. Mulligan, D. Carmean, G. Seelig, L. Ceze, K. Strauss, Random access in large-scale DNA data storage. *Nat. Biotechnol.* **36**, 242–248 (2018).

6. L. Ceze, J. Nivala, K. Strauss, Molecular digital data storage using DNA. *Nature Reviews Genetics 2019 20:8* **20**, 456–466 (2019).

7. P. L. Antkowiak, J. Lietard, M. Z. Darestani, M. M. Somoza, W. J. Stark, R. Heckel, R. N. Grass, Low cost DNA data storage using photolithographic synthesis and advanced information reconstruction and error correction. *Nat. Commun.* **11**, 5345- (2020).

8. L. Žemaitis, R. Palepšienė, S. Juzėnas, G. Alzbutas, P. Y. Burgi, T. Heinis, J. Charmet, S. A. Suter, M. Jost, R. Raišutis, F. Simmel, I. Galminas, High-performance protocol for ultra-short DNA sequencing using Oxford Nanopore Technology (ONT). *PLoS One* **20**, e0318040 (2025).

9. Y. J. Chen, C. N. Takahashi, L. Organick, C. Bee, S. D. Ang, P. Weiss, B. Peck, G. Seelig, L. Ceze, K. Strauss, Quantifying molecular bias in DNA data storage. *Nature Communications 2020 11:1* **11**, 3264- (2020).

10. W. H. Press, J. A. Hawkins, J. M. Schaub, J. M. Schaub, I. J. Finkelstein, HEDGES error-correcting code for DNA storage corrects indels and allows sequence constraints. *Proc. Natl. Acad. Sci. U. S. A.* **117**, 18489–18496 (2020).

11. K. B. Mullis, The unusual origin of the polymerase chain reaction. *Sci. Am.* **262** (1990).

12. R. K. Saiki, S. Scharf, F. Faloona, K. B. Mullis, G. T. Horn, H. A. Erlich, N. Arnheim, Enzymatic amplification of β-globin genomic sequences and restriction site analysis for diagnosis of sickle cell anemia. *Science (1979).* **230** (1985).

13. J. Cheng, E. Zhang, R. Sun, K. Zhang, F. Zhang, J. Zhao, S. Feng, B. Liu, Implementation of Rapid Nucleic Acid Amplification Based on the Super Large Thermoelectric Cooler Rapid Temperature Rise and Fall Heating Module. *Biosensors 2024, Vol. 14, Page 379* **14**, 379 (2024).

14. C. T. Wittwer, G. C. Fillmore, D. R. Hillyard, Automated polymerase chain reaction in capillary tubes with hot air. *Nucleic Acids Res.* **17** (1989).

15. C. D. Ahrberg, A. Manz, B. G. Chung, Polymerase chain reaction in microfluidic devices. *Lab Chip* **16** (2016).

16. Y. Zhang, P. Ozdemir, Microfluidic DNA amplification-A review. *Anal. Chim. Acta* **638**, 115–125 (2009).

17. J. Charmet, P. Arosio, T. P. J. Knowles, Microfluidics for Protein Biophysics. *J. Mol. Biol.* **430**, 565–580 (2018).

18. J. F. Huggett, C. A. Foy, V. Benes, K. Emslie, J. A. Garson, R. Haynes, J. Hellemans, M. Kubista, R. D. Mueller, T. Nolan, M. W. Pfaffl, G. L. Shipley, J. Vandesompele, C. T. Wittwer, S. A. Bustin, The digital MIQE guidelines: Minimum information for publication of quantitative digital PCR experiments. *Clin. Chem.* **59** (2013).

19. B. J. Hindson, K. D. Ness, D. A. Masquelier, P. Belgrader, N. J. Heredia, A. J. Makarewicz, I. J. Bright, M. Y. Lucero, A. L. Hiddessen, T. C. Legler, T. K. Kitano, M. R. Hodel, J. F. Petersen, P. W. Wyatt, E. R. Steenblock, P. H. Shah, L. J. Bousse, C. B. Troup, J. C. Mellen, D. K. Wittmann, N. G. Erndt, T. H. Cauley, R. T. Koehler, A. P. So, S. Dube, K. A. Rose, L. Montesclaros, S. Wang, D. P. Stumbo, S. P. Hodges, S. Romine, F. P. Milanovich, H. E. White, J. F. Regan, G. A. Karlin-Neumann, C. M. Hindson, S. Saxonov, B. W. Colston, High-throughput droplet digital PCR system for absolute quantitation of DNA copy number. *Anal. Chem.* **83** (2011).

20. A. S. Basu, Digital Assays Part I: Partitioning Statistics and Digital PCR. *SLAS Technol.* **22**, 369–386 (2017).

21. G. M. Whitesides, The origins and the future of microfluidics. *Nature* **442**, 368–373 (2006).

22. T. M. Squires, S. R. Quake, Microfluidics: Fluid physics at the nanoliter scale. *Rev. Mod. Phys.* **77** (2005).

23. OT-2 Liquid Handler | Opentrons Lab Automation from $10,000 | Opentrons. https://opentrons.com/robots/ot-2#resources.

24. B. Bajar, E. Wang, S. Zhang, M. Lin, J. Chu, A Guide to Fluorescent Protein FRET Pairs. *Sensors* **16**, 1488 (2016).

25. P. C. Gach, K. Iwai, P. W. Kim, N. J. Hillson, A. K. Singh, Droplet microfluidics for synthetic biology. *Lab Chip* **17**, 3388–3400 (2017).

26. J. Kalová, R. Mareš, The temperature dependence of the surface tension of water. *AIP Conf. Proc.* **2047**, 20007 (2018).

27. S. W. Mayer, Dependence of Surface Tension on Temperature. *J. Chem. Phys.* **38**, 1803–1808 (1963).

28. J. Kalová, R. Mareš, Temperature Dependence of the Surface Tension of Water, Including the Supercooled Region. *International Journal of Thermophysics 2022 43:10* **43**, 154- (2022).

29. L. Korson, W. Drost-Hansen, F. J. Millero, Viscosity of water at various temperatures. *Journal of Physical Chemistry* **73**, 34–39 (2002).

30. J. Aslanzadeh, Brief Review: Preventing PCR Amplification Carryover Contamination in a Clinical Laboratory. *Ann. Clin. Lab. Sci.* **43**, 389–396 (2004).

31. V. Seitz, S. Schaper, A. Dröge, D. Lenze, M. Hummel, S. Hennig, A new method to prevent carry-over contaminations in two-step PCR NGS library preparations. *Nucleic Acids Res.* **43**, e135–e135 (2015).

32. L. Erlandsson, L. P. Nielsen, A. Fomsgaard, Amp-PCR: Combining a Random Unbiased Phi29-Amplification with a Specific Real-Time PCR, Performed in One Tube to Increase PCR Sensitivity. *PLoS One* **5**, e15719 (2010).

33. DNAMIC: DNA Microfactory for Data Archiving | EU Project. https://dnamic.org/.

34. OLOS: The Swiss solution for managing research data. https://olos.swiss/.

## Acknowledgements

We thank Frédéric Flahaut for support on injection moulding, and Duygucan Gül and Sarah Scotton for support on molecular biology experiments. We thank Friedrich Simmel for his contribution to funding acquisition.

## Funding

This research was funded by the European Innovation Council Pathfinder Program, under the project DNAMIC (DNA Microfactory for Autonomous Archiving), Grant Agreement No. 101115389 and Swiss State Secretariat for Education, Research and Innovation (SERI) under contracts number 23.00300 and 23.00487.

## Authors Contributions

VB, JG: Investigation, Formal Analysis, Data Curation, Validation and Methodology. SJ, KK, LŽ, DK, IG: Investigation, Methodology, Validation, Data curation. PYB: Investigation, Methodology, Validation. SW, VR, SA, PN: Investigation, Methodology. BVDS, AC: Resources, Data curation. MJ: Resources, Funding Acquisition. RR, TH: Funding Acquisition. JC: Conceptualisation, Supervision, Formal analysis, Validation, Writing – original draft, Funding acquisition. All authors: Writing – review & editing.

## Competing interest.

The authors confirm no competing interests.

## Data and materials availability.

The codec for DNA data encoding/decoding is available in Gitlab: https://gitlab.unige.ch/dlcm/community/dlcm-backend/. The code used for DNA data analysis is available in Github: https://github.com/genomika-lt/PCRobot. Raw sequencing data was uploaded to ENA under PRJEB111897 accession number.

# **Supplementary Materials:** High-fidelity robotic PCR amplification

Vincent Beguin[1,*], Jean Grétillat[2,*], Kornelija Kaminskaitė[3], Simonas Juzenas[3], Dainius Kirsnauskas [3], Pierre-Yves Burgi[4], Samuel Wenger[1], Valentin Remonnay[1], Silvia Angeloni[1], Patrick Neuenschwander[1], Bart van der Schoot[5], Augustin Cerveaux[5], Thomas Heinis[6], Renaldas Raisutis[7], Martin Jost[8], Lukas Zemaitis[3], Ignas Galminas[3#], Jérôme Charmet[1,2,9#]

10. *School of Engineering−HE-Arc Ingénierie, HES-SO University of Applied Sciences Western Switzerland, Neuchâtel, Switzerland*
11. *School of Precision and Biomedical Engineering, University of Bern, Bern, Switzerland*
12. *Department of DNA data storage, Genomika, Kaunas, Lithuania*
13. *IT Department, University of Geneva, Geneva, Switzerland*
14. *Seyonic SA, Neuchâtel, Switzerland*
15. *Department of Computing, Imperial College London, London, United Kingdom*
16. *Ultrasound Research Institute, Kaunas University of Technology, Kaunas, Lithuania*
17. *MABEAL GmbH, Graz, Austria*
18. *Warwick Medical School, University of Warwick, Coventry, United Kingdom*

* the authors contributed equally to the work.

[#] corresponding authors

**3-bath PCRobot.**

Early implementations of the PCRobot relied on 3-bath container heated to the desired temperature plateaus, using hot plates (Figure **S1**). The Seyonic pipettor single head, attached to a 3-axis positioning system (OMNI UNIVERSAL ROBOT by Tecan, internal reference n°0103), moves from one bath to the next to perform PCR into the tip. The thermal response along with the position of the robot are shown in Figure **S1a**. Later we favoured the simple PCRobot implementation with a single bath and a robotic arm, as detailed in the manuscript. Early measurements were compared and validated using the MiniOne Kit (thermocycler and gel electrophoresis). The PCR amplification measurement was performed using the MiniOne dye added to the mix before dispensing on the gel for electrophoresis. Control experiments were performed using the MiniOne thermocycler (same conditions) followed by the same gel electrophoresis. Figure **S1c** shows the results that correspond to a 20-cycle process performed in 22 minutes for the PCRobot and in 15 minutes for the MiniOne thermocycler.

**Impulse response and thermal expansion**

The impulse responses of our tips were evaluated using filtered distilled water (25 µL). Figure **S5a)** and **b)** show heating and cooling respectively. Considering the response of 1[st] order system, $T(t) = T_1 + (T_0 - T_1)(e^{-t/\tau})$, the time constant is measured and the theorical model of the impulse response can be solved. Resulting time constants for 4 scenarios are shown in Table **S1**.

When matter changes temperature, it changes shape, area, volume, and density in response. This tendency is called thermal expansion. The polypropylene tip containing the two air columns that trap the PCR solution represented by water will expand. The upper end of the tip is closed by the valve and the lower end is open, Figure **S5c**.

In the general case of a gas, liquid, or solid, the volumetric coefficient of thermal expansion is given by:

$$\alpha = \alpha_V = \frac{1}{V}\left(\frac{\partial V}{\partial T}\right)_p \quad [1]$$

The subscript $p$ indicates constant pressure. For linear dimensions we can rewrite:

$$V_{final} = V_{initial}\left(1 + \alpha_V\left(T_{final} - T_{initial}\right)\right) \quad [2]$$

Using a linear expansion coefficient for air from Granta EduPack level3: $\alpha_{L\,air} = 3330\,\mu\frac{1}{°C}$, the volume of expanded air is calculate to $\Delta V_A = 162\mu L$ and $\Delta V_A = 174\mu L$ for 50 µL and 175 µL tips respectively. These values should not in theory result in any liquid being ejected from the tip, based on experimental air gaps measurements. Therefore, thermal expansion alone cannot explain the liquid loss.

**Encoding**

DNAMIC logo in svg format was packed into OLOS package and encoded with 30% error correction redundancy. Package size is 8.1KB leading to 352 oligos inside the encoded design file.

**CODEC Description.**

A DNA-encoded binary file is represented by a list of DNA sequences, called oligonucleotides, with A,T,G, and C sequences of 243 letters on each line making up a text file. Each sequence contains forward and reverse primers of 18 nucleotides (nts) length used for PCR amplification and, in-between, 207 nts for data (Fig. **S9a**). The file identity is mapped to specific forward and reverse primers, randomly generated by sampling a hash function (SHA256) and selected on the basis of Gibbs free energy to optimize PCR, and minimal free energy to avoid secondary structures, and separated by a Hamming distance greater than 10 to limit hybridization between primers. Furthermore, to ensure efficient binding of both primers and avoid asymmetric amplification, the difference between the melting temperatures of the forward and reverse primers must be kept below 2 degrees Celcius.

The data contains error correction codes (inner code), a portion of the file (payload), and an index to identify which part of the file has been encoded. Encoding and decoding are carried out using symbols. A symbol is equivalent to 16 bits encoded by 9 nucleotides (Fig. **S9b**). Input files are treated as byte sequences, transformed into an *n* by *m* matrix, where *n* (number of rows), which corresponds to the payload, is equal to 34 (equivalent to 153 nts), and *m* (number of columns) is equal to the rounded result of dividing the length of the file byte sequence by 34, to ensure that all data fits into the matrix. If the last column is not filled, it is completed with dummy data, and if the number of columns is odd, an extra column is added with dummy data. To avoid repeating patterns, dummy data are defined by incrementing a counter modulo 256. The index is coded on 4 bytes, making it possible to encode up to a 146 GB file.

Reed–Solomon codes are applied to detect and correct multiple symbol errors for matrix columns (inner codes) and matrix rows (outer codes). The inner code, represented by 8 bytes, is added to the first row of the matrix (see Figure **S9c**). The outer code is calculated for each row joining two bytes (which is why an extra column is added if the number is odd). Since an RS symbol is equivalent to 16 bits, the maximum number of bytes that can be used to represent columns is 131070 bytes. Beyond this size, the matrix must be separated into blocks.

Considering the target redundancy *r*, which depends on the error rate resulting from the various DNA processes (i.e. synthesis, preservation, PCR, and sequencing), the number $e_c$ corresponding to the outer code columns added to each block is given by

$$e_c = \frac{r \cdot \min(m, 131070)}{1 - r}$$

with $e_c$ rounded to even. On the basis of sequencing experiments using nanopore technology, we set *r* in the 10-30% range. The minimum (rounded up) of $n_b$ blocks required to store all information (excluding the outer code) is therefore

$$n_b = \frac{m}{131070 - e_c}$$

Blocks can contain different amounts of data, and, to distribute symbols evenly, the average number of bytes required per block is

$$\overline{m} = \frac{m}{n_b}$$

with $\overline{m}$ rounded to even.

The 2-byte conversion is achieved by mapping three Galois Field (GF) elements of size 41 by base conversion ($256^2$ to $41^3$). Individual columns are converted to DNA using the GF(41) codon wheel shown in Figure 1B, ensuring that no base is repeated more than three times.

**Primer** sequences used for amplification are ACGAGCGTGAGTACTCTG for forward primer and CGCCAGCTCCAGTCGGTG for reverse primer.

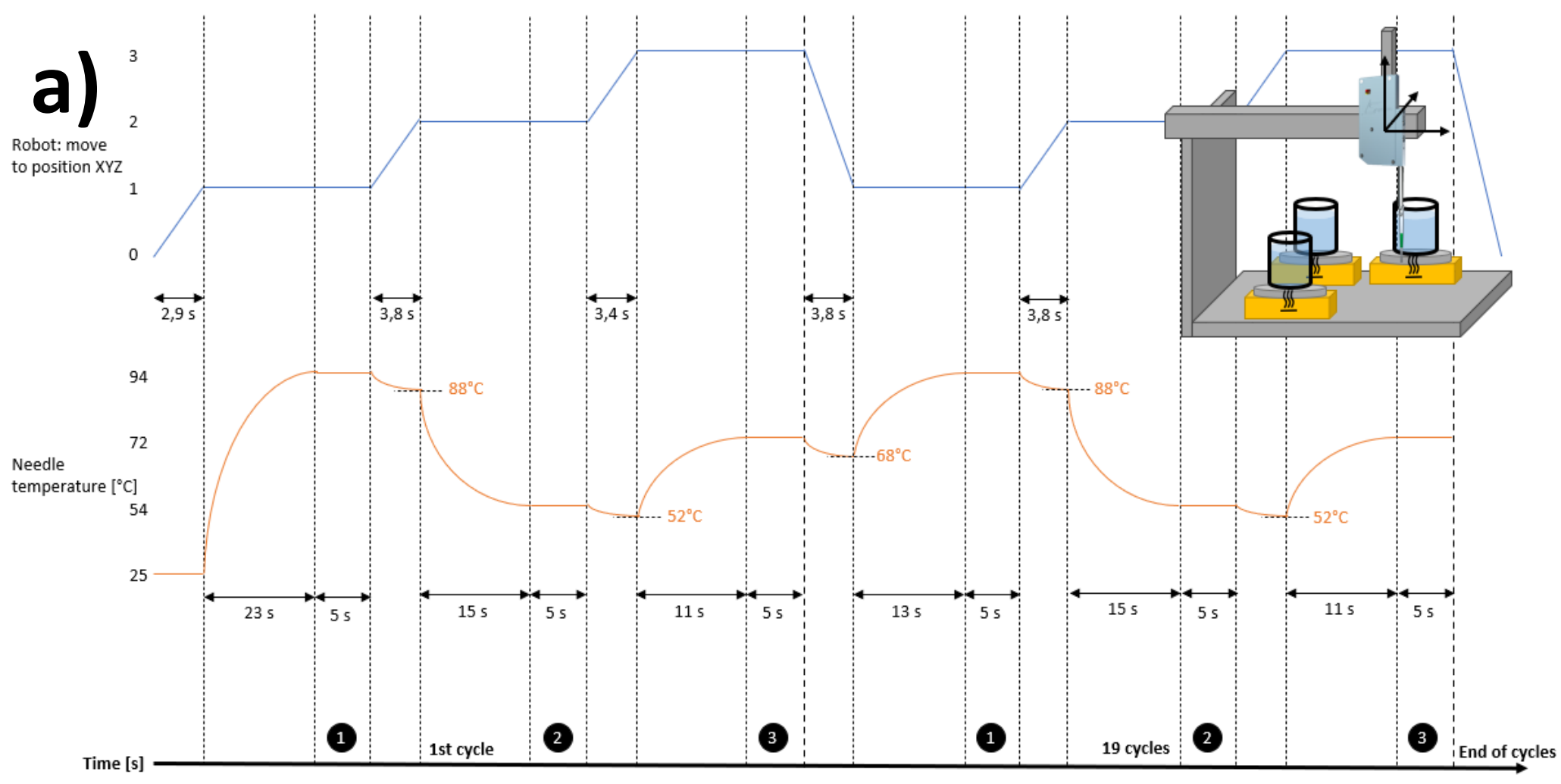

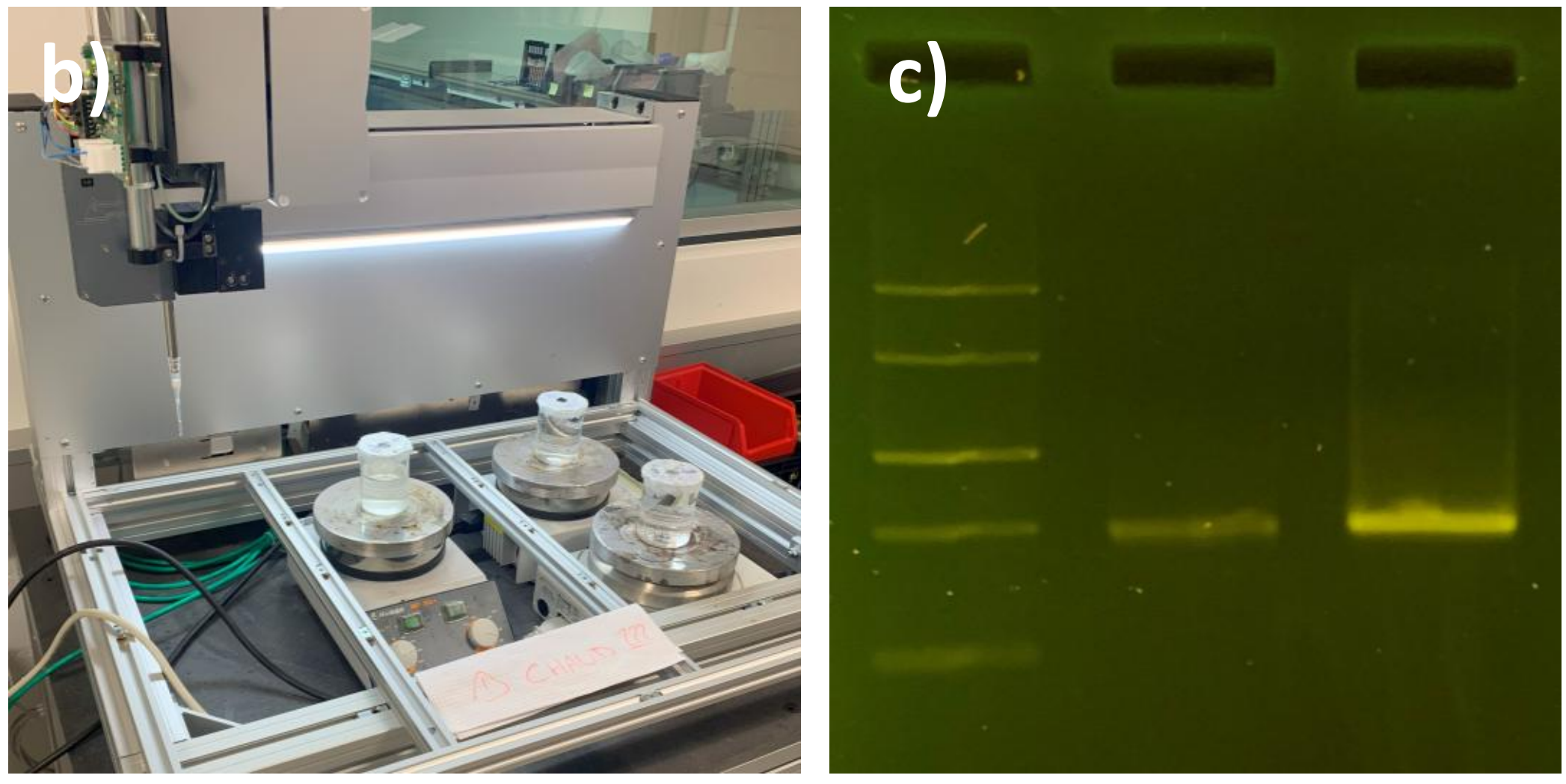

*__Figure S1.__ Three bath robotic-enabled DNA amplification set-up. a) Temperature profile and corresponding robot position. Inset shows the principle. b) image of the set-up. c) Electropherogram (MiniOne kit) comparing the performance of the MiniOne thermocycler with the 3 bath PCRobot device. Lane 1 - ladder, lanes 2 – MiniOne thermocycler: three replicate samples amplified using the MiniOne thermocycler; lanes 3 – PCRobot. The band visible across the lane corresponds to the expected PCR product (amplicon 304 nt).*

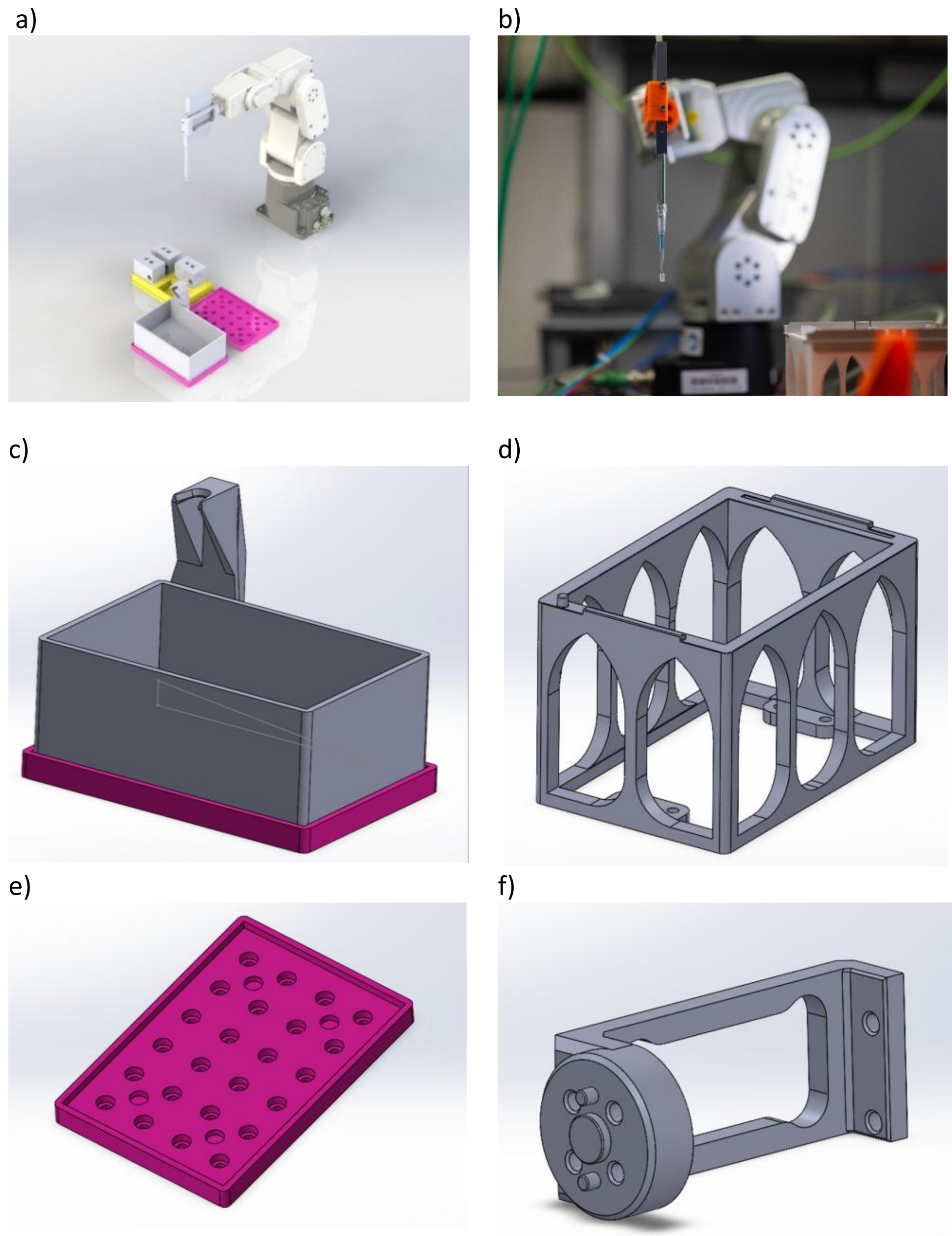

***Figure S2.*** *PCRobot and components. a) schemcatic overview of the original 3 bath set-up. b) picture of the PCRobot with emphasis on the pipettor and the tip, c) waste tip container, d) pipette tip holder, e) normalised – well plate format – base tip rack and waste container, f) pipettor holder.*

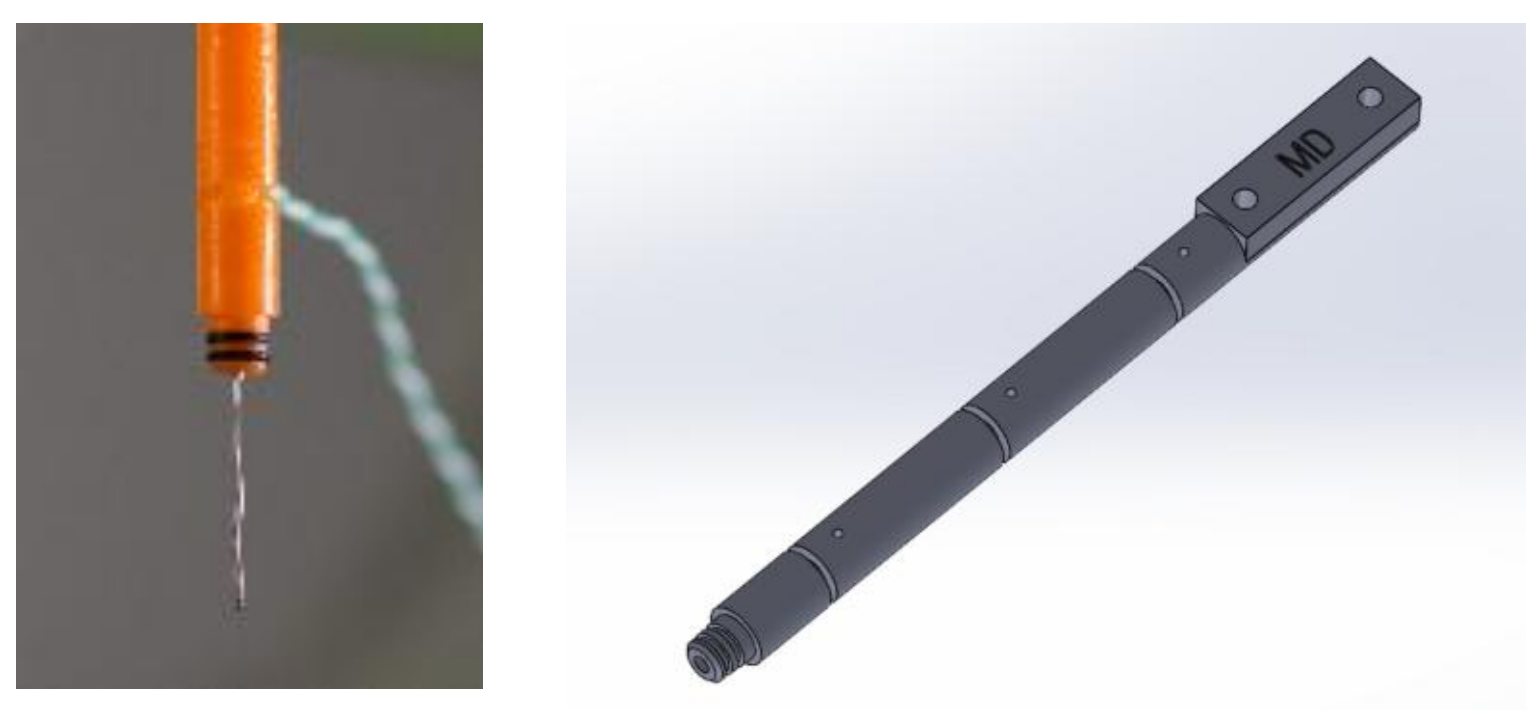

***Figure S3.*** *Custom-made thermocouple and mandrel (photograph and CAD file)*

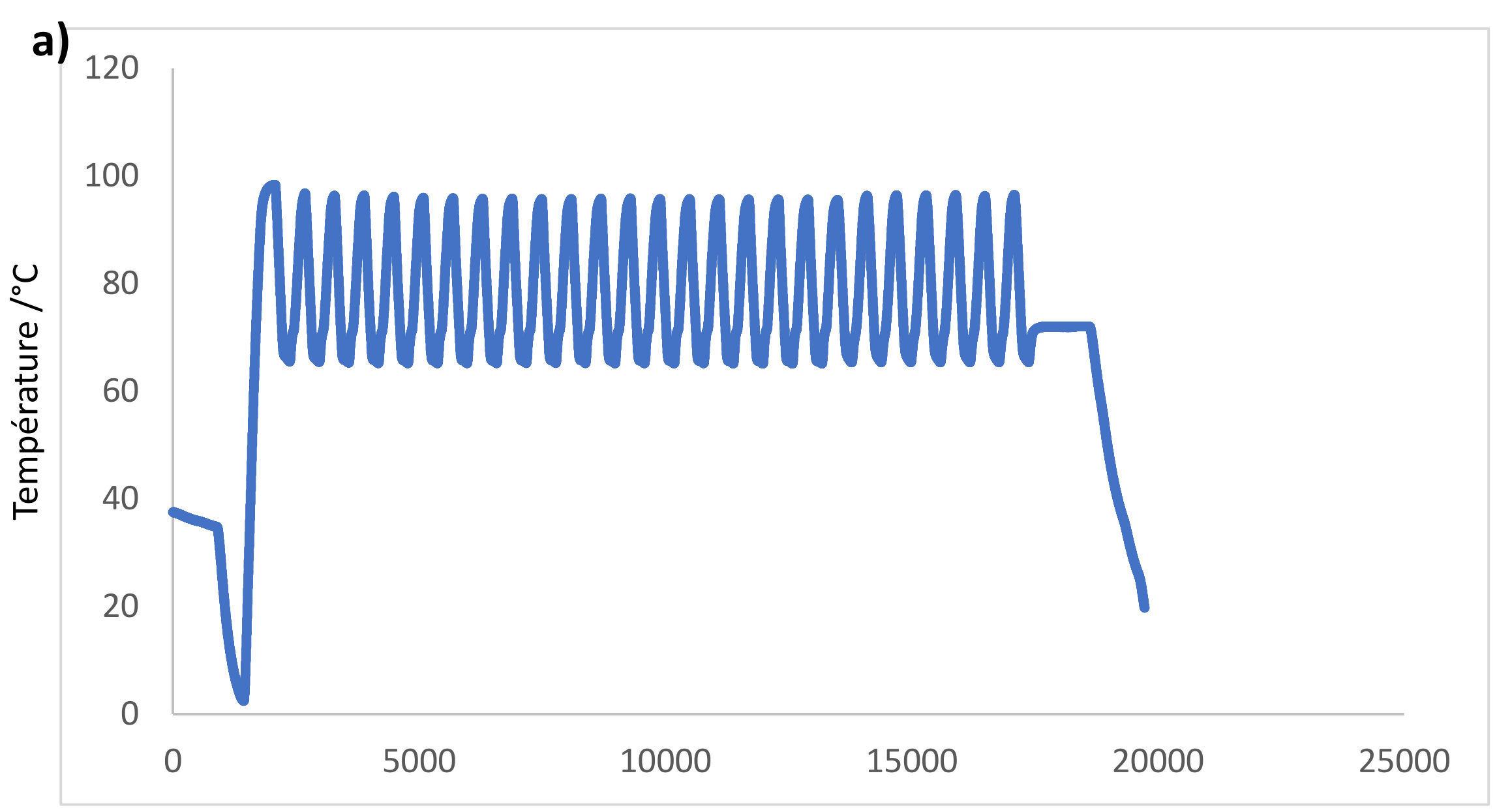

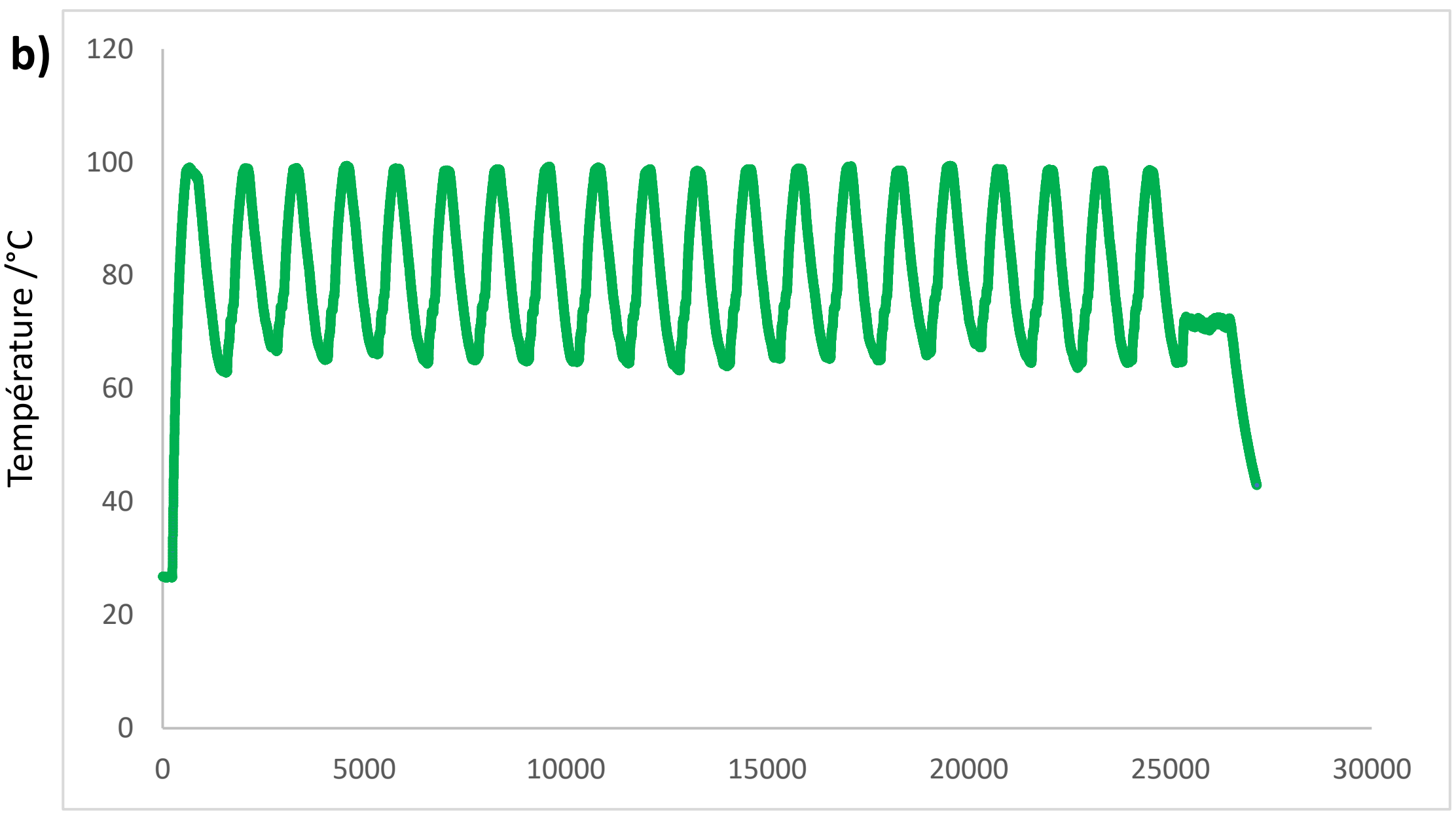

***Figure S4.*** *Thermal response of the Bio-Gener (Labgene) thermocycler (a) and the PCRobot (b), as a function of the scan number (logger scanning frequency). Thermal cycling consisted of an initial denaturation at 98 °C for approximately 30 s, followed by cycles including brief denaturation (≈10 s at 98 °C), annealing (≈15 s at 65 °C), and short extension (≈5 s at 72 °C), with a final elongation step of ~2 min at 72 °C.*

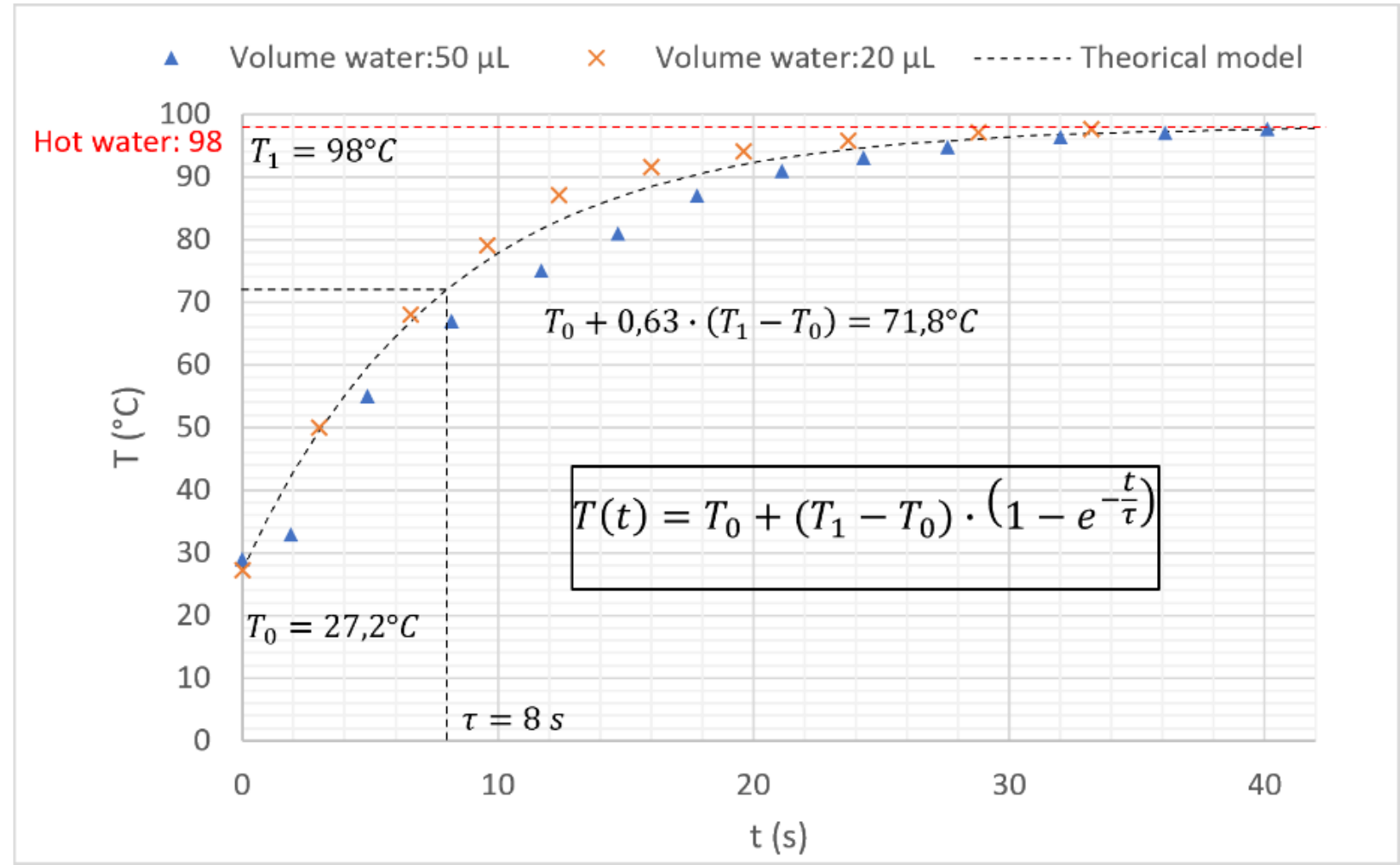

b)

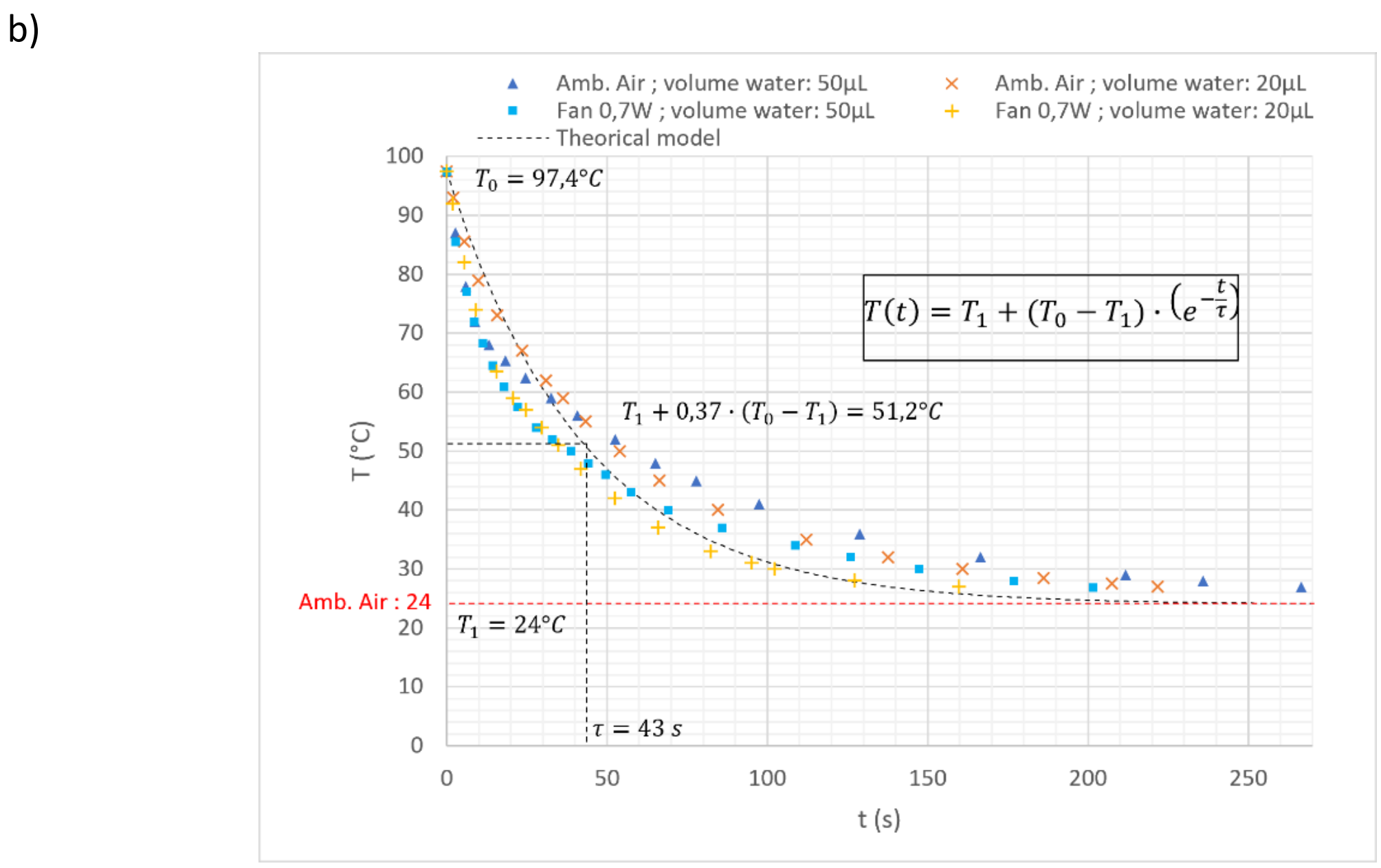

c)

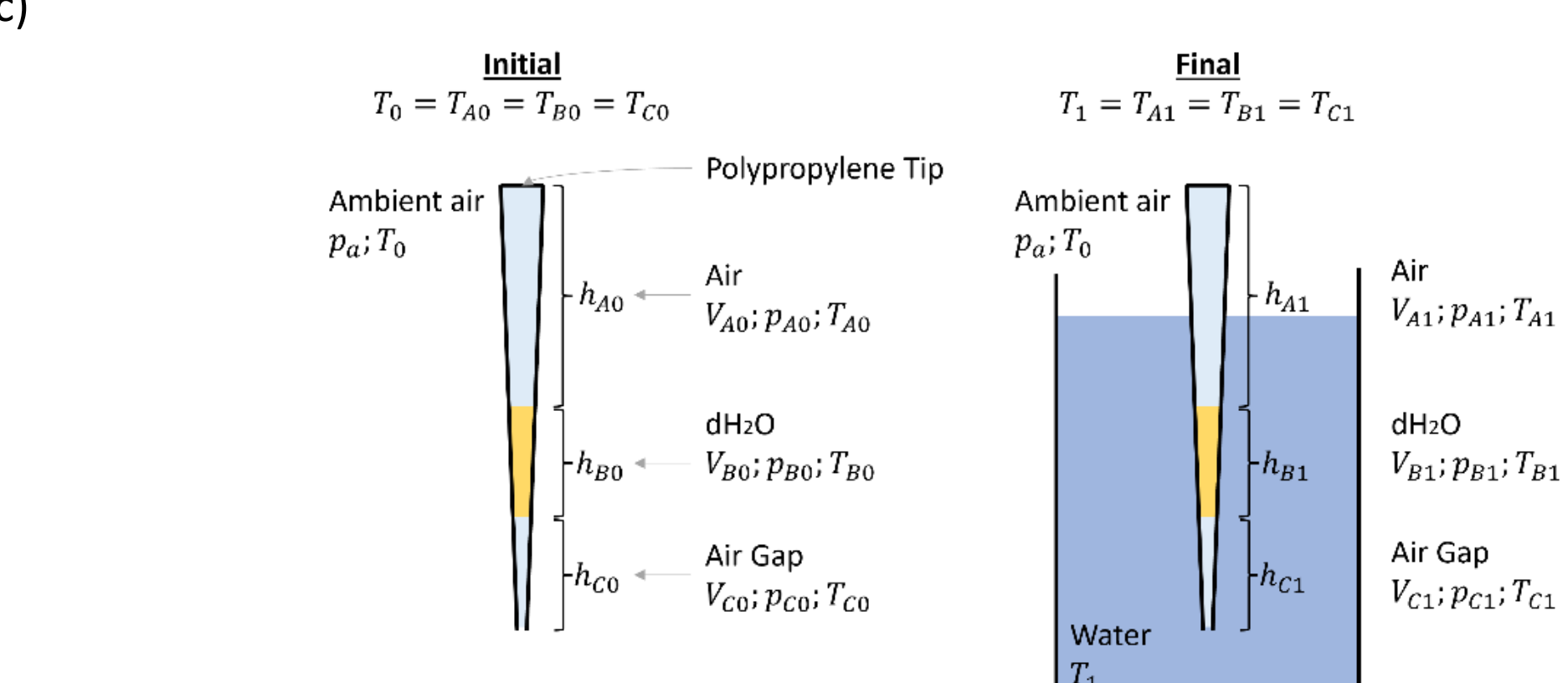

***Figure S5***. *Measure of time constant of the impulse response at a unit step of (a) 98°C (boiling water) and (b) 24°C (ambient air). c) Schematics of thermal expansion of system*

**NTC and amplicon**

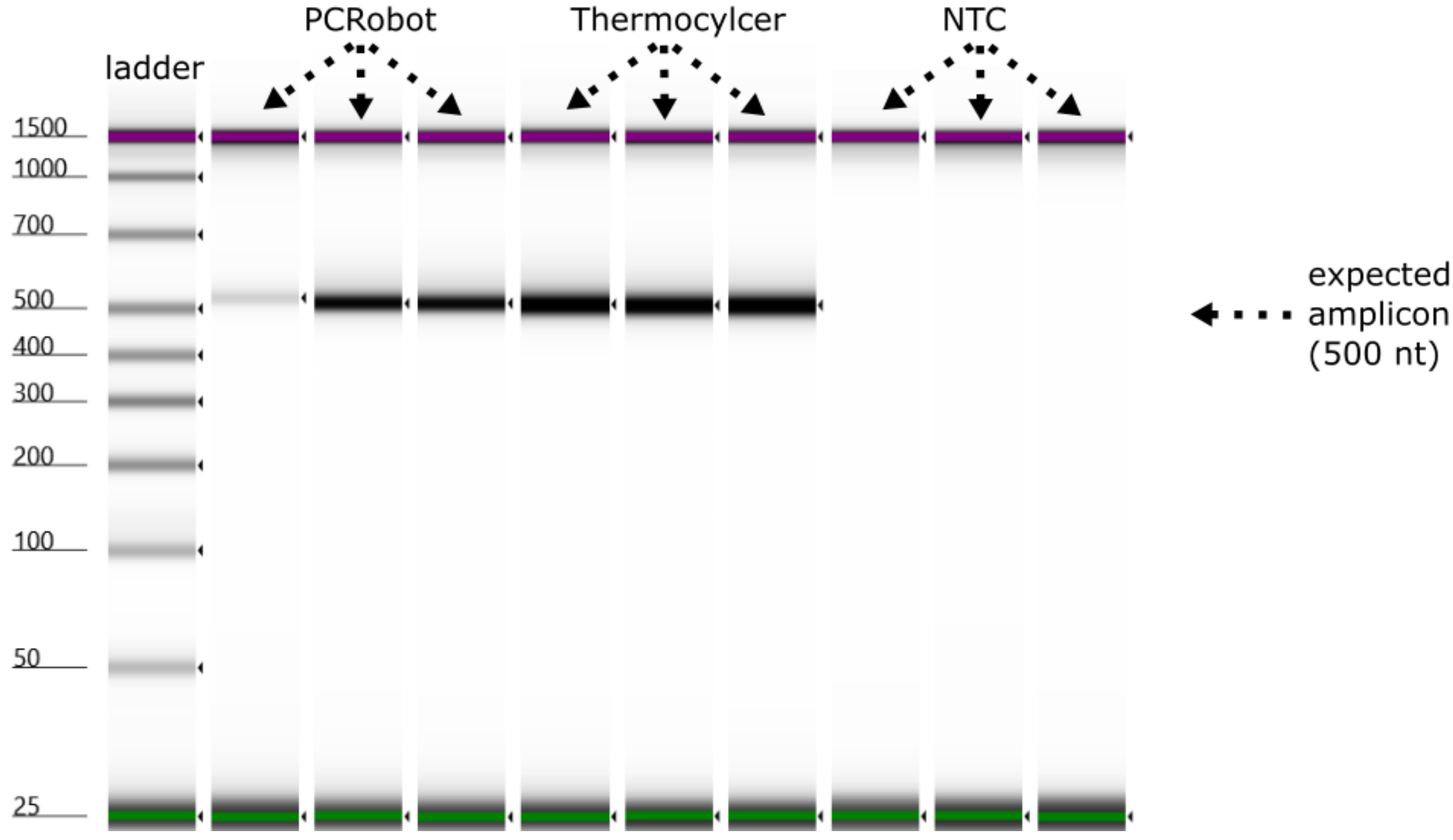

***Figure S7.*** *Electropherogram (Agilent 2200 TapeStation, D1000 ScreenTape assay) comparing the performance of the PCRobot, the thermocycler and a no template control (NTC). Lane 1 - ladder, lanes 2-4 - PCRobot: three replicate samples amplified using the PCRobot. Lane 2 is an outlier due to wrong manipulation resulting in partial sample loss; lanes 5-7 - Thermocylcer: three replicate samples amplified using the thermocylcer. The band visible across all lanes (2-7) corresponds to the expected PCR product (amplicon). Its location aligns with the 500 bp size marker on the ladder, confirming successful amplification in all six reactions. Lanes 7-9 three replicate samples amplified using No Template Control (NTC) with 1 ng of template DNA in the oil bath. No amplification observed, confirming that no contamination occurred.*

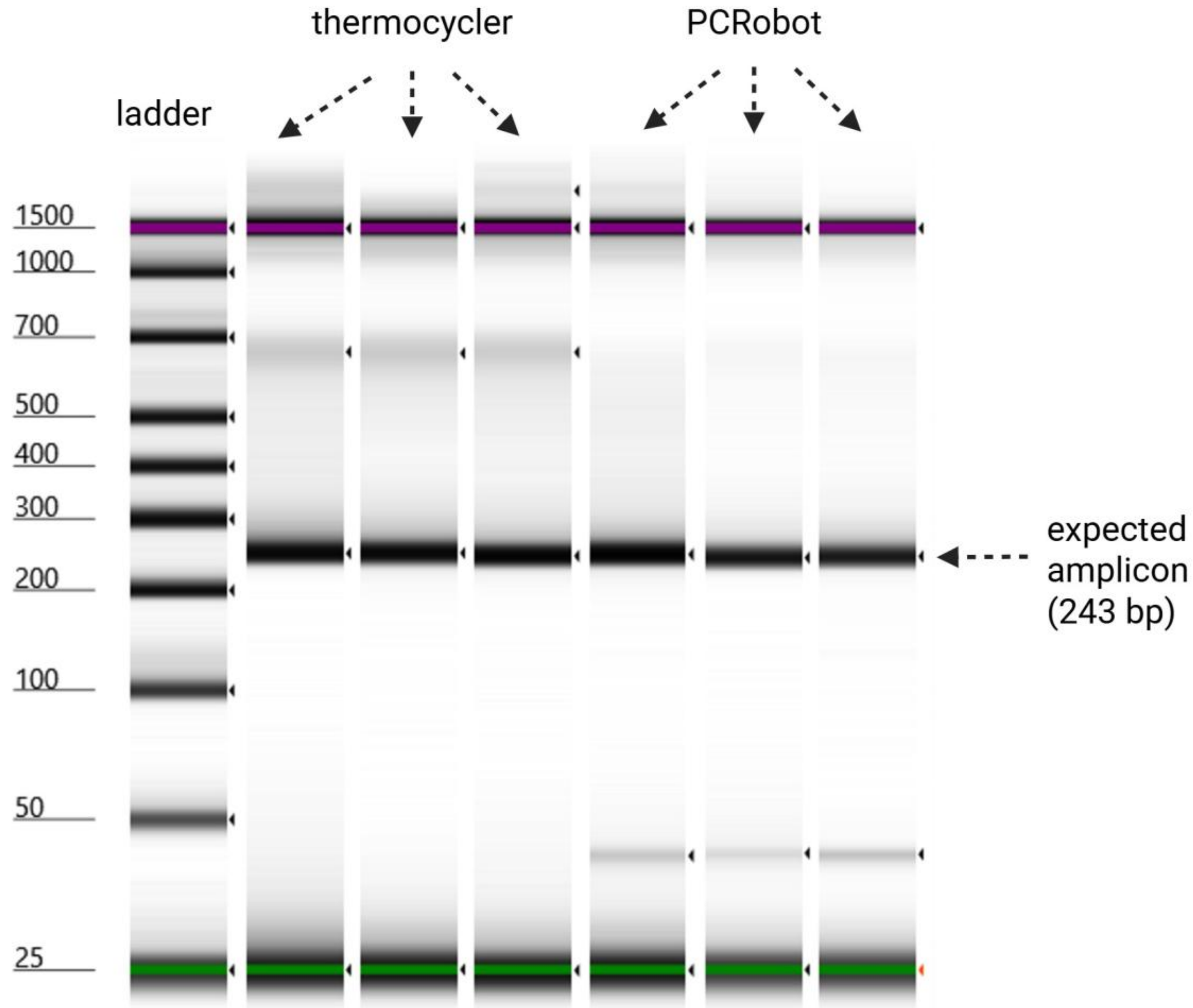

***Figure S8.*** *Electropherogram* (Agilent 2200 TapeStation, D1000 ScreenTape assay) *comparing the performance of a Thermocycler with the PCRobot device. Lane 1 - ladder, lanes 2-4 - thermocycler: three replicate samples amplified using a Thermocycler; lanes 5-7 - PCRobot: three replicate samples amplified using the PCRobot device. The band visible across all lanes (2-7) corresponds to the expected PCR product (amplicon). Its location aligns with the 243 bp size marker on the ladder, confirming successful amplification in all six reactions.*

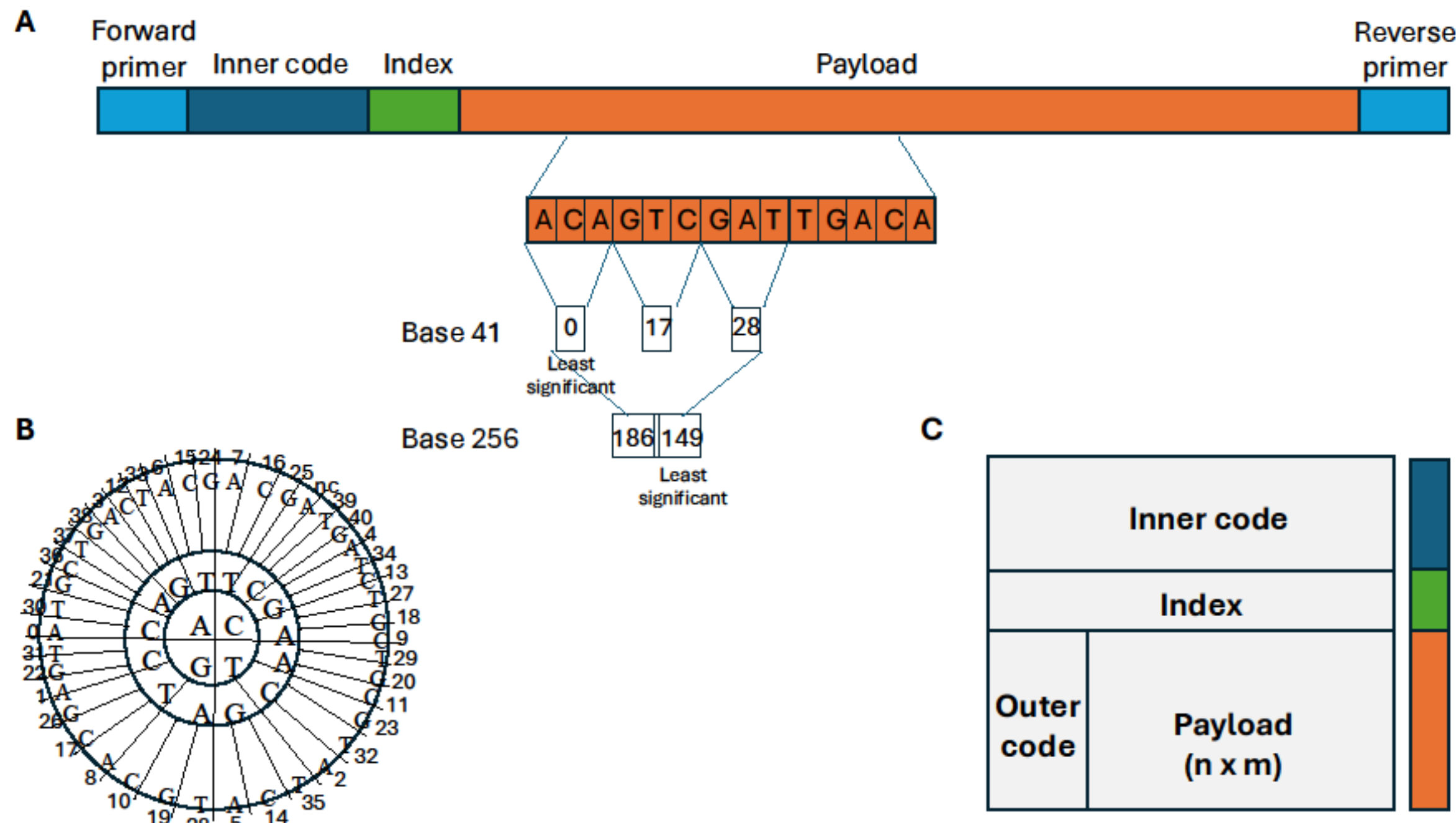

***Figure S9***. *A) Details of the oligonucleotide structure, with an illustration of a 2-byte-to-nucleotides conversion; B) GF(41) is derived from the codon wheel; C) General block structure.*

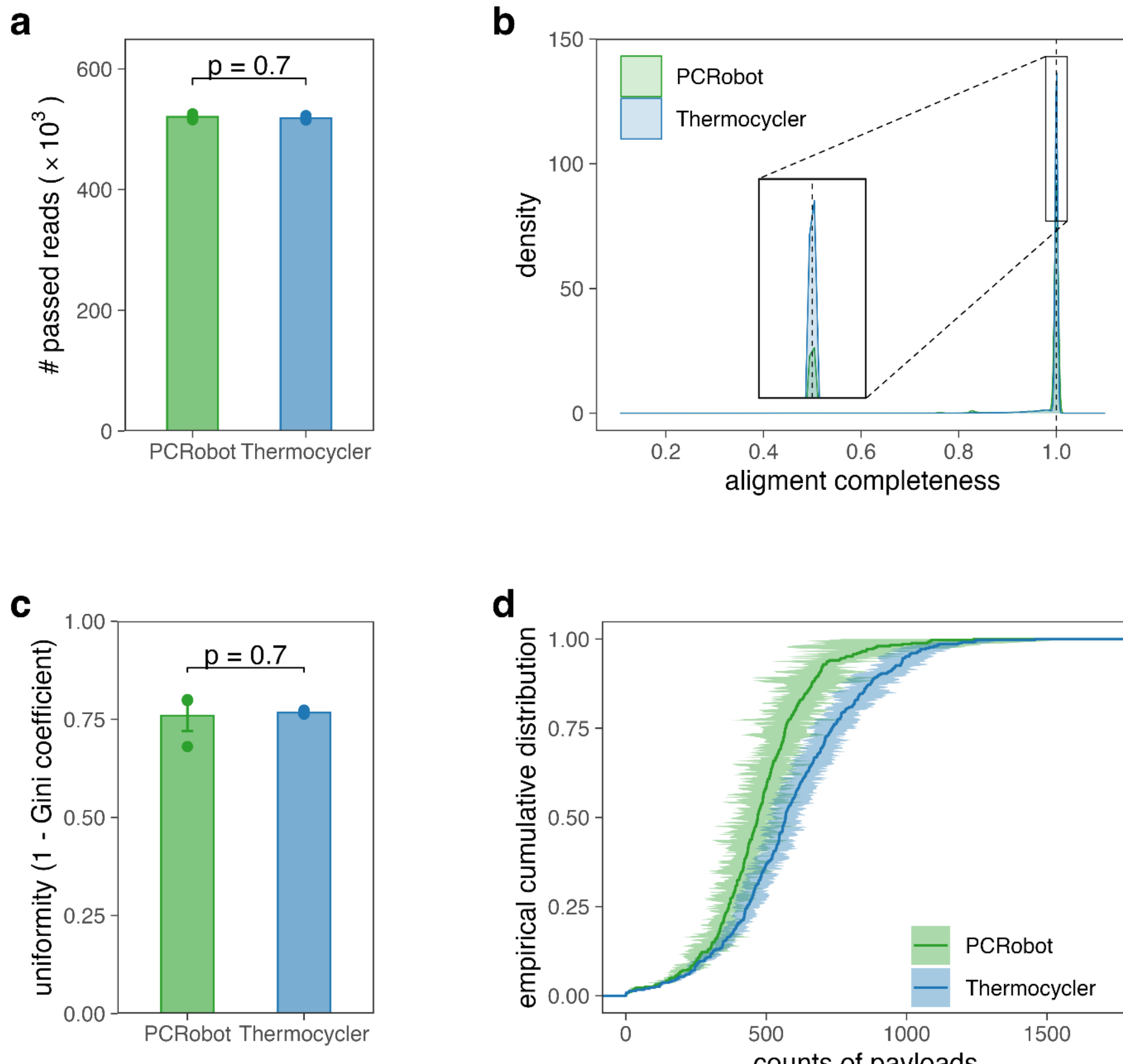

***Figure 10.*** *Performance comparison of PCRobot and Thermocycler for oligopool PCR amplification and sequencing in the context of DNA data storage. (**a**) Number of reads passing quality filtering (qs > 9) for each device. Bars represent mean ± SEM across samples (n = individual samples shown as points). Statistical significance assessed by Wilcoxon test. (**b**) Density distribution of alignment completeness (fraction of read length aligned to reference) for mapped reads from both devices. Dashed vertical line indicates complete alignment. Inset shows magnified view of the peak region near perfect alignment. (**c**) Total number of reads mapped to payload sequences (allowing max four mismatches). Bars represent mean ± SEM across samples (n = individual samples shown as points). Statistical significance assessed by Wilcoxon test. (**c**) Uniformity of payload coverage calculated as 1 minus the Gini coefficient, where higher values indicate more uniform distribution of reads across targets. Bars represent mean ± SEM across samples (n = individual samples shown as points). Statistical significance assessed by Wilcoxon test. (**d**) Empirical cumulative distribution functions (ECDF) of mean read counts per payload. Step lines show the ECDF with shaded ribbons representing ± 1 SEM. Higher curves indicate greater representation of payloads at a given count threshold. All data represent biological replicates with PCRobot (blue) and Thermocycler (green) conditions. P-values in panels **a**, and **c** are from two-sided Wilcoxon rank-sum tests.*

**Table. S1 Impulse responses**

| Impulse response of 20µL $dH_2O$ in 200µL tip: | | $\tau\ [s]$ |
|---|---|---|
| [a] | immersed in water at 98°C (heating). | 8 |
| [b] | immersed in water at 76°C (heating). | 6.2 |
| [c] | immersed in air at 24°C (cooling). | 43 |
| [d] | immersed in water at 54°C (cooling). | 6 |